\documentclass[twocolumn]{aastex63}
\UseRawInputEncoding
\usepackage{color,comment}
\usepackage[normalem]{ulem}
\usepackage[caption=false]{subfig}
\usepackage{graphicx}
\usepackage{verbatim}
\usepackage{mathtools} 
\usepackage{float}
\usepackage{wrapfig}
\usepackage{appendix}

\received{}
\revised{}
\accepted{}

\shorttitle{Origin of High-z DSFGs}
\shortauthors{S. Lower et al.}

\begin{document}

\title{Cosmic Sands: The Origin of Dusty, Star-forming Galaxies in the Epoch of Reionization}

\author[0000-0003-4422-8595]{Sidney Lower}
\affil{Department of Astronomy, University of Florida, 211 Bryant Space Science Center, Gainesville, FL, 32611, USA}
\author[0000-0002-7064-4309]{Desika Narayanan}
\affil{Department of Astronomy, University of Florida, 211 Bryant Space Science Center, Gainesville, FL, 32611, USA}
\affil{University of Florida Informatics Institute, 432 Newell Drive, CISE Bldg E251 Gainesville, FL, 32611, US}
\affil{Cosmic Dawn Centre at the Niels Bohr Institue, University of Copenhagen and DTU-Space, Technical University of Denmark}
\author[0000-0001-8015-2298]{Qi Li}
\affil{Max-Planck-Institut f\"{u}r Astrophysik, Karl-Schwarzschild-Str. 1, Garching b. M\"{u}nchen D-85741, Germany}
\author[0000-0003-2842-9434]{Romeel Dav{\'{e}}}
\affil{Institute for Astronomy, Royal Observatory, University of Edinburgh, Edinburgh, EH9 3HJ, UK}
\affil{Department of Physics and Astronomy, University of the Western Cape, Bellville, 7535, South Africa}

\begin{abstract}

We present the Cosmic Sands suite of cosmological zoom-in simulations based on the {\sc simba} galaxy formation model in order to study the build up of the first massive and dusty galaxies in the early Universe. Residing in the most massive halos, we find that the compact proto-massive galaxies undergo nearly continuous mergers with smaller subhalos, boosting star formation rates (SFRs) and the build up of stellar mass. The galaxies are already appreciably chemically evolved by $z=10$, with modeled dust masses comparable to those inferred from observations in the same epoch. We track gas accretion onto the galaxies to understand how extreme SFRs can be sustained by these early systems. We find that smooth gas accretion can maintain SFRs above $250 \: \mathrm{M}_{\odot}/\mathrm{yr}$ but to achieve SFRs that boost galaxies well above the main sequence, a larger perturbation like a gas-rich major merger is necessary to trigger a starburst episode. Post-processing the Cosmic Sands simulations with dust radiative transfer, we find that while the infrared luminosities of the most dust rich galaxies are comparable to local ULIRGs, they are substantially dimmer than classical $z=2$ sub-millimeter galaxies. We end with a discussion on the possible reasons for this discrepancy at the highest masses and the future work we intend to carry out to study the chemical enrichment of the earliest dusty galaxies.  

\end{abstract}

\section{Introduction}

By the time the Universe was 1 billion years old, the dawn of galaxy formation was already well underway. The processes that govern the formation of the most massive galaxies during this time are still relatively uncertain. From a theoretical standpoint, studying these processes requires connecting physics across several orders of magnitude in spatial (as well as mass and temporal) scales, from star formation on sub-parsec scales to galactic outflows and the baryon cycle on kilo-parsec scales to galaxy-galaxy interactions over mega-parsec scales. Modeling the formation of early massive galaxies has been a numerical challenge for several decades. The first attempts with semi-analytic models focused on explaining the origin of high redshift submillimeter galaxies (SMGs), Lyman-break galaxies (LBGs), and quasars \citep{guiderdoni_1998_smg_sam, devriend_guiderdoni_2000_smg_sam,somerville_lbg_sam, granato_2004_quasar_model, baugh_2005_simulating_smgs, swinbank_2008_smg_sam} individually. Later, more robust galaxy formation frameworks were implemented that evolved these populations together \citep{fontanot_2007_sam, somerville_2012_sam,bethermin_2017_smg_sam, lacey_2016_galform, lagos_2019_shark_sam, triani_2020_dustysage, hutter_2021_sam} and used hydrodynamic models coupled with dust radiative transfer to produce high-z populations of massive galaxies with realistic growth histories and radiative properties \citep{dave_2010_smg_hydro,narayanan_2010_smgs_hydro_a, narayanan_2010_smgs_hydro_b, shimizu_2012_smg_hydro, hayward_2013_smg, narayanan_2015_smgs_nature,poole_2016_dark_ages_hydro, mcalpine_2019_eagle_smgs,kim_2019_highz_quasars_hydro, lovell_2021_flares, lovell_2021_simba_counts}.

\looseness=-1
Of particular interest is the population of infrared luminous SMGs that have been detected as early as redshift $z=6.9$ \citep{marrone_2018_spt0311}. These systems are some of the most extreme in the Universe, with stellar masses estimated to be comparable to the Milky Way that formed in less than 1 billion years after the Big Bang. Their intense infrared luminosities, comparable to those of local ULIRG galaxies, are thought to be the result of massive dust reservoirs heated by compact starbursts. The rarity of these systems make them difficult to study, with number counts estimated to be $<10 / \mathrm{deg}^2$ past $z=6$ \citep{wardlow_2011_less_counts, simpson_2014_aless, ivison_2016_dsfg_counts, zavala_2021_mora, reuter_2020_spt}. Given their rareness, they are not perfectly traced by dark-matter over-densities \citep{chapman_2009_smgs_dont_trace, miller_2015_smgs_dont_trace}. Even with their intrinsic rarity, these extreme systems were the first insights into early galaxy formation, challenging theoretical models that attempted to explain their rapid build-up and stellar and dust masses. 

For instance the onset and nature of star formation and the impact of local and global feedback mechanisms remain relatively uncertain. Similarly, the question of how these same galaxies build up dust reservoirs in excess of $10^7 \: \mathrm{M}_{\odot}$ is also difficult to reconcile with current theory. Estimates for dust yields from Type II supernovae (SNe) can explain the dust masses for galaxies at $z > 6$ only if the destruction rates are negligible \citep{michalowski_2015_dust, burgarella_2020_early_dust}, as chemical enrichment from evolved asymptotic branch (AGB) stars is negligible at these redshifts.

With the additional constraints of ``normal" dusty, star-forming galaxies at $z>5-6$ from ALMA large program (e.g., REBELs \citealt{bouwens_rebels}, ALPINE \citealt{lefevre_2020_alpine}), we are beginning to understand that even relatively small galaxies have sizeable dust mass reservoirs and that chemical enrichment is not just a characteristic of the most extreme systems. Establishing how and when the build up of both stellar and dust masses occurs is necessary to understand the origin of the first infrared luminous galaxies. 

\looseness=-1
Two recent works have significantly advanced our understanding of early galaxy formation in a cosmological context: the FirstLight project \citep{ceverino_2017_first_light}, a large suite of hydrodynamic zoom-in simulations, successfully reproduced observed galaxy stellar mass and ultra-violet (UV) luminosity functions and the star-forming main sequence relation out to redshift $z=10$. The FLARES project, based on {\sc eagle} physics \citep{crain_2015_eagle, schaller_2015_eagle, schaye_2015_eagle, mcalpine_2016_eagle_cat} and implementing a novel simulation scheme to model over-densities in the Epoch of Reionization (EoR), has also successfully formed galaxies with physical and photometric properties consistent with observations out to redshift $z=10$ \citep{lovell_2021_flares, vijayan_2021_flares_2}. This said, neither the FirstLight nor FLARES projects are able to produce galaxies during the Epoch of Reionization that span of the range of colors and masses inferred by current constraints, especially the most extreme systems observed to date \citep[e.g.][]{reuter_2020_spt, watson_2015_A1689, marrone_2018_spt0311, hashimoto_2019_b14, endsley_2022_SB_AGN}.

\looseness=-1
Aiming to place infrared bright galaxies into the context of galaxy evolution, \cite{narayanan_2015_smgs_nature} presented a theory for the evolution of SMGs at intermediate redshifts $(1<z<2)$. These galaxies experience gas rich minor mergers with small subhalos that trigger bursts of star formation, resulting in a compact stellar core with extended long-lived infrared emission.  \cite{lovell_2021_simba_counts} extended this framework into a cosmological context by nearly reproducing observed SMG number counts with the {\sc simba} hydrodynamical simulation, echoing the findings from \cite{narayanan_2015_smgs_nature} that infrared luminous phases are a natural consequence of massive galaxy evolution. However, it is not obvious if this framework extends out to high redshift $(z>6)$. Indeed, neither \citet{narayanan_2015_smgs_nature} nor \citet{lovell_2021_simba_counts} modeled a sufficiently large volume with high enough resolution to form dusty star forming galaxies at $z>6$.

\looseness=-1
To this end, we present the Cosmic Sands sample of dusty galaxies, occupying an effective volume of $(0.5\times 100 \: \mathrm{Mpc})^{3}$ and modeled with the {\sc simba} galaxy formation suite \citep{dave_simba} that includes well-constrained models for star formation, stellar feedback, black hole growth and active galactic nuclei (AGN) feedback, and dust production growth, and destruction. These high resolution zoom-in simulations are the first to couple stellar evolution and dust growth and enable detailed studies into the formation pathways of early massive galaxies, with a focus on modeling the most extreme star-forming galaxies in the Epoch of Reionization. In this paper, we aim to understand the mechanisms through which dusty, star-forming galaxies (DSFGs) build up their extensive amounts of gas, stellar, and dust masses in the first billion years of cosmic history, focusing on their formation pathways and the drivers of their early star formation.

\looseness=-1
In what follows, we outline the galaxy formation model {\sc simba} and the post-processing radiative transfer code {\sc powderday} in Section \S \ref{sec:methods}, the physical properties of the galaxies and comparisons to observations from the literature in Section \S \ref{sec:overview}, and answer the question ``What sustains the intense SFRs observed in high-z DSFGs?" in Section \S \ref{sec:gas_accretion}. Finally, in Section \S \ref{sec:discussion}, we discuss to what degree our Cosmic Sands galaxies match the observed properties of high-z galaxies as well as the fundamental uncertainties in our model.

\section{Numerical Methods}\label{sec:methods}

\subsection{The {\sc simba} Galaxy Formation Model}

\looseness=-1
Our massive galaxy sample is generated from the {\sc simba} galaxy formation model. {\sc Simba} \citep{dave_simba} is based on the {\sc Gizmo} gravity and hydrodynamics code \citep{hopkins_2015_gizmo} and includes models describing heating and cooling, star formation, chemical enrichment, feedback from stellar winds, dust production and growth, and blackhole (BH) accretion and feedback. We briefly summarize the key aspects of each model. 

\looseness=-1
Star formation occurs in dense molecular clouds, with rates governed by the density of H$_2$ divided by the local dynamical timescale. The H$_2$ fraction is modeled with the sub-grid prescription of \citet{krumholz_2011_h2} depending on the gas-phase metallicity and local gas column density. We impose a minimum density limit $(n_\mathrm{H} \geq 0.13 \: \mathrm{cm}^{-3})$ below which stars do not form. 

\begin{figure}[t]
    \centering
    \includegraphics[width=0.40\textwidth]{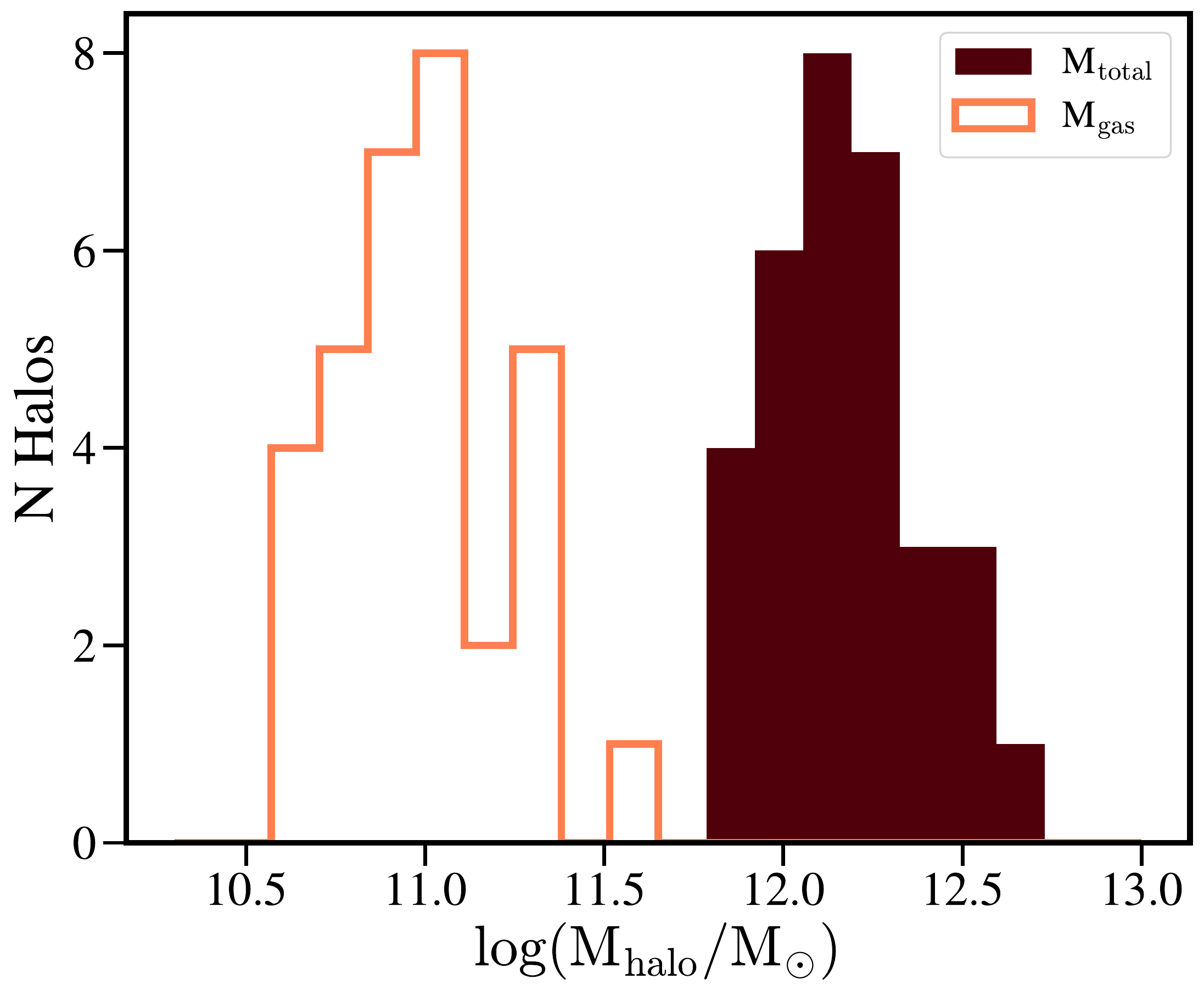}\\
    \includegraphics[width=0.40\textwidth]{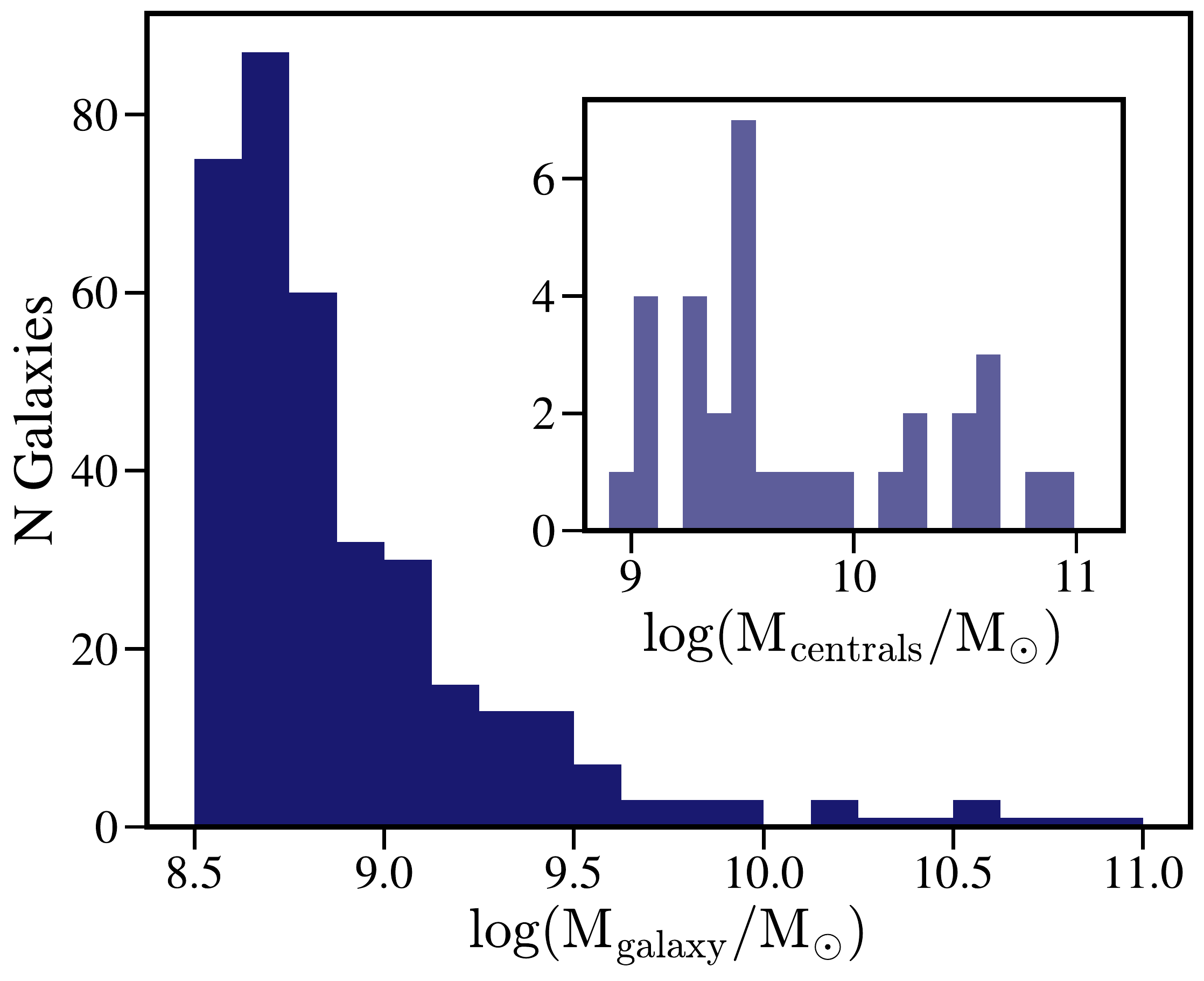}\\
    \includegraphics[width=0.40\textwidth]{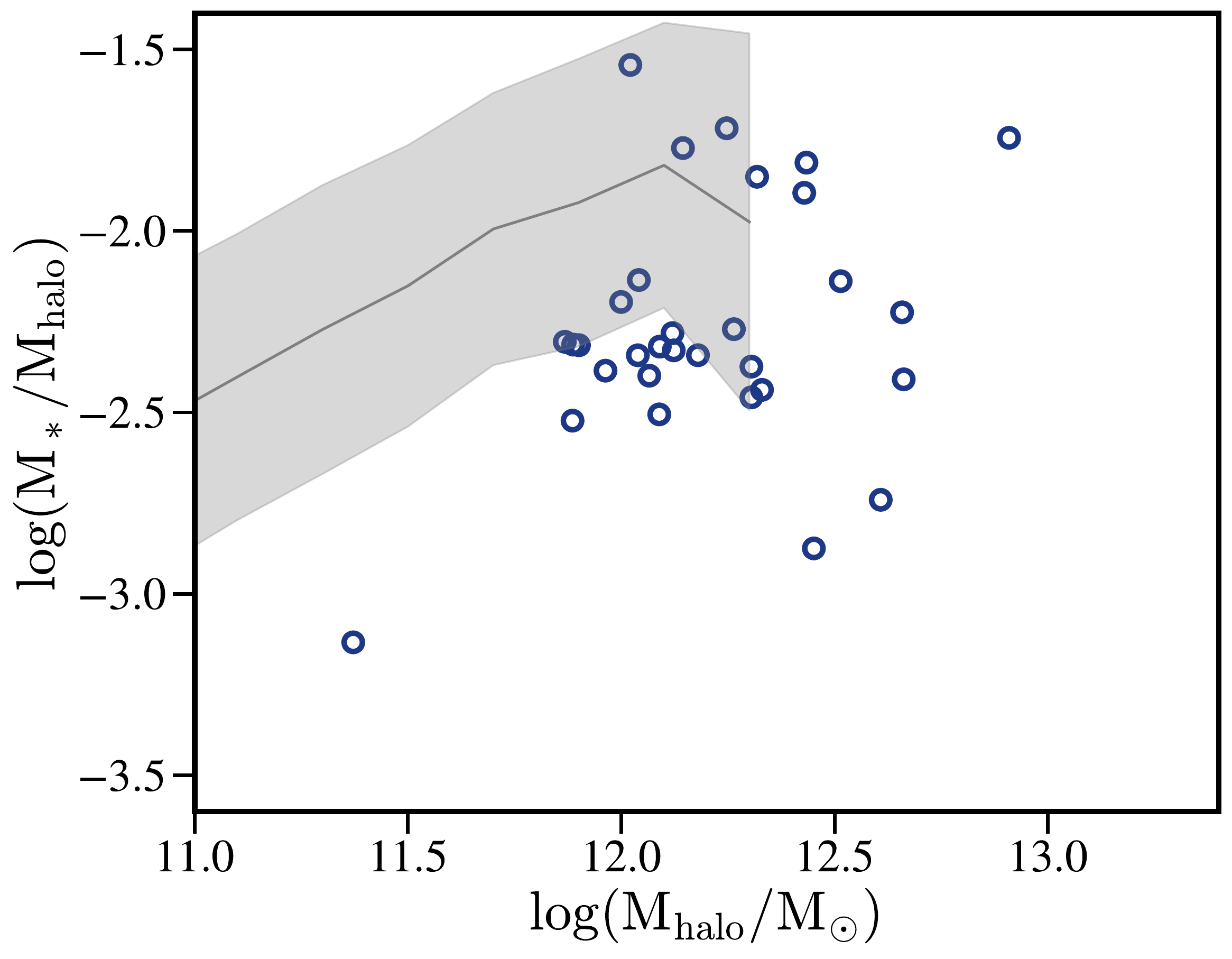}
    \caption{\textbf{Top}: Redshift $z = 6.7$ Cosmic Sands halo mass distribution. The filled in histogram shows the distribution of total (gas + dark matter + stars) halo mass while the outlined histogram shows only the halo gas mass distribution. \textbf{Middle}: Mass (gas + stars) distribution for all galaxies residing in the halos in the top panel. The inset plot highlights the central galaxies, defined as the galaxy residing in the minimum potential well. \textbf{Bottom}: M$_*$ - M$_\mathrm{halo}$ relation at $z=6.7$, where the stellar mass is the sum of over all subhalos and the halo mass is the total (gas + dark matter + stellar) mass. The gray curve and region shows the relation from the Universe Machine \citep{behroozi_2019_umachine}.}
    \label{fig:mass_dist}
\end{figure}

\looseness=-1
{\sc Simba} uses the {\sc grackle-3} library \citep{smith_2017_grackle} to model radiative cooling and photoionization heating including a self-consistent model for self-shielding based on \citet{rahmati_2013_self_shield}. The chemical enrichment model tracks $11$ elements from from Type Ia and II supernovae (SNe) and asymptotic giant branch (AGB) stars with yields following \cite{nomoto_simba_sne_yields}, \cite{iwamoto_1999_sne_yields}, and \cite{Oppenheimer_2006_agb_yields}, respectively. 

\looseness=-1
Stellar feedback is implemented with contributions from Type II SNe, radiation pressure, and stellar winds. The two-component stellar winds adopt the mass-loading factor scaling from {\sc fire} \citep{angles-alcazar_FIRE} with wind velocities given by \cite{muratov_2015}. Metal-enriched winds extract metals from nearby particles to represent the local enrichment by the SNe driving the wind. Feedback via active galactic nuclei (AGN) is implemented as a two-phase jet (high accretion rate ) and radiative (low accretion rate) model. Thermal energy is injected into the surrounding interstellar medium (ISM) at high accretion rates, while BH-driven winds are produced at low accretion rates. 
\looseness=-1

Dust is modeled self-consistently, and on the fly in the galaxy evolution simulations. Following the fiducial models of \cite{qi_dust}, dust is produced by the condensation of metals ejected from Type II SNe and AGB stars, and is allowed to grow and erode depending on local ISM temperature, density, and shocks from Type I and Type II SNe. Dust production from stellar sources occurs with fixed condensation efficiencies following the models of \citet{ferrarotti_2006_agb_dust} for AGB stars and \citet{bianchi_2007_sne_dust} for SNe. Dust growth in the ISM occurs via accretion of metals onto the seeded dust grains; accretion rates are governed by the local gas density, temperature, and metallicity, with the reference density scaled by the ratio of the median ISM density in the zooms to the median ISM density in the lower resolution 100 Mpc/h {\sc simba} box \citep{qi_dust}.

Lastly, dust grains are eroded via thermal sputtering and can be destroyed entirely by SNe shocks. The sputtering timescale depends on the gas density and temperature, with the sputtering rate flattening above temperatures of $2\times10^6 \: \mathrm{K}$. SNe shocks are not directly modeled, thus we implement a SNe destruction mode with a timescale dependent on the mass of gas shocked to velocities of at least $100 \mathrm{km}/\mathrm{s}$ with a fixed grain destruction efficiency of $0.3$ following \citet{mckee_1989_shock}. In hot winds, during star formation, and in any gas that is impacted by AGN feedback, dust is assumed to be completely destroyed.

\subsection{Zoom In Technique}

\looseness=-1
We generate our suite of massive galaxies using a zoom-in technique that enables our simulations to be generated at higher mass resolutions, and thus finer detail, while still maintaining the cosmological scale of a larger box. We do this by creating 32 dark-matter only simulations with initial conditions set by {\sc music} \citep{hahn_abel_2011_music}. Each simulation is 25 Mpc$^3$ in volume and evolved from $z=299$ to $z=0$. We then select the largest halo from each box at redshift $z = 2$ using {\sc caesar} \citep{caesar} and construct an ellipsoidal mask around all particles within $2.5\times$ the radius of the maximum distance (i.e. the farthest away) dark matter particle in the halo. This is defined as the Lagrangian high-resolution region to be re-simulated at a higher resolution with baryon physics included. The zoom-ins reach an effective mass resolution of m$_\mathrm{baryon} \sim 2.9\times10^5$ M$_{\odot}$ for particles within the zoom-in radius. We identify galaxies within each halo again using {\sc caesar} with a 6-D friends-of-friends galaxy finder based on the number of bound stellar particles in a system (a minimum of 32 stellar particles defines a galaxy). 

\looseness=-1
Snapshots for the zoom-ins are saved in intervals of 30 Myr beginning at redshift $z_\mathrm{init} = 10$ with a select subsample of halos saved at earlier times ($z_{init} = 25$) and at finer time resolution ($\Delta t = 15$ Myr). In total, our 32 zoom in simulations represent an effective volume of 4$\times$50 Mpc$^3$ or 0.5$\times$ a 100 Mpc$^3$ box. We note, however, that the use of several $25$ Mpc/$h$ boxes will not capture the largest perturbations and may be missing some of the larger scale structure that would be present in a $100$ Mpc/$h$ box, thus slightly biasing our sample against the highest mass halos.

\subsection{3D Dust Radiative Transfer}\label{sec:pd}

\looseness=-1
We couple our hydrodynamic zoom-in simulations with 3D dust radiative transfer (RT) to generate UV-FIR SEDs for each massive galaxy. We use the radiative transfer code {\sc powderday}\footnote{https://github.com/dnarayanan/powderday} \citep{pd_2020} to construct the synthetic SEDs by first generating with {\sc fsps} \citep{fsps_1, fsps_2} the dust free SEDs for the star particles within each cell using the stellar ages and metallicities as returned from the cosmological simulations. For these, we assume a \cite{kroupa_initial_2002} stellar IMF and the {\sc mist} stellar isochrones \citep{mist_1, mist_2}. These {\sc fsps} stellar SEDs are then propagated through the dusty ISM. The diffuse dust content is derived from the on-the-fly self-consistent model of \cite{qi_dust}. As {\sc simba} contains a 'passive' dust model, with a single grain size ($0.1\mu$m), the dust is assumed to have extinction properties following the carbonaceous and silicate mix of \cite{draine_infrared_2007}. This model includes the \cite{weingartner_draine} size distribution and the \cite{draine_03} renormalization relative to hydrogen. We assume $R_{\rm V} \equiv A_{\rm V}/E(B-V) = 3.15$.

\looseness=-1
Polycyclic aromatic hydrocarbons (PAHs) are included following the \citet{robitaille_pahs} model in which PAHs are assumed to occupy a constant fraction of the dust mass (here, modeled as grains with size $a<20$ \AA) and occupying $5.86\%$ of the dust mass. The dust emissivities follow the \citet{draine_infrared_2007} model, though are parameterized in terms of the mean intensity absorbed by grains, rather than the average interstellar radiation field as in the original Draine \& Li model. 

\looseness=-1
The radiative transfer propagates through the dusty ISM in a Monte Carlo fashion using {\sc hyperion} \citep{robitaille_2011_hyperion}, which follows the \cite{lucy_rt} algorithm in order to determine the equilibrium dust temperature in each cell. We iterate until the energy absorbed by $99\%$ of the cells has changed by less than $1\%$. 

\begin{figure*}[t]
    \centering
    \includegraphics[width=0.9\textwidth]{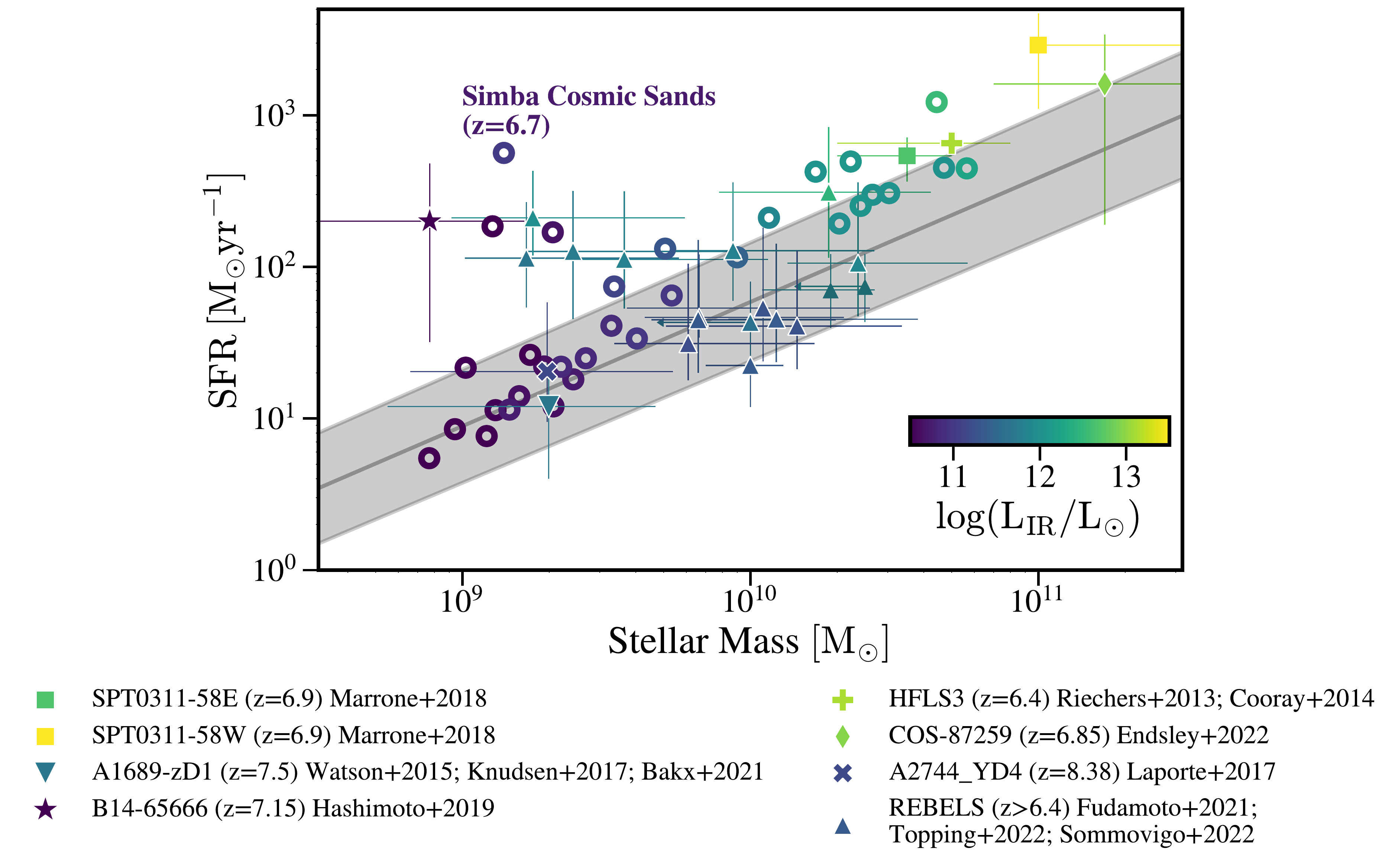}
    \caption{SFR as a function of stellar mass for our zoom-in simulations (colored circles) and observations (various colored symbols) at redshift $z = 6.4-7.5$. The simulated galaxies show reasonable consistency with the observations across a wide range of stellar masses. Symbols are colored by the total infrared luminosity. The gray region shows the star-forming main sequence relation at redshift $z=6.7$ as measured by \cite{speagle_2014_sfms}. Note that for galaxy COS-87259, L$_{IR}$ measurements exist for both an AGN component and thermal dust emission, where the latter value is plotted above.}
    \label{fig:sfr_mass}
\end{figure*}

\looseness=-1
We perform RT on snapshots for one viewing angle with a fixed aperture radius of $50 \: \mathrm{kpc}$, roughly equal to the beam size of the SCUBA-2 camera on JCMT at $850 \: \mu m$ for $z > 2$ \citep{geach_2017_scuba}, following the methodology of \citep{lovell_2021_simba_counts}. This aperture is generally much larger than the optical or FIR size of an individual galaxies, meaning the RT SEDs include contributions from satellites and neighboring galaxies. Because of this, the physical properties of the galaxies we report in our analysis are for all particles within this aperture, not just those associated with a single FOF-identified galaxy.

\section{Cosmic Sands: Massive, Dusty Galaxies at Cosmic Dawn} \label{sec:overview}

\subsection{Sample Overview \& Observational Constraints}

\looseness=-1
In Figure \ref{fig:mass_dist}, we plot the distribution of masses for the halos and galaxies at redshift $z = 6.7$ as well as the stellar mass - halo mass relation. The halos span an order of magnitude in gas masses, which drives the diversity in central galaxy masses; as we'll discuss below the gas accretion and star formation rates are largely set by the halo gas mass alone. While we derive the initial conditions for our zoom-in simulations from a series of dark matter only simulations that, in aggregate, encompass an effective volume of  4$\times$50 Mpc$^3$, in practice our models are not necessarily representative of this volume since we pick only the most massive halo within each sub-volume to zoom in on.

\looseness=-1
By redshift $6.7$, 9 halos have produced central galaxies with a stellar mass above $10^{10}$ M$_{\odot}$. We choose this epoch to compare our simulated massive galaxies with observations of high-z galaxies around the same redshift ($z > 6.4$) as we are primarily interested in the first billion years of evolution. These observations include galaxies from the REBELS program \citep{bouwens_rebels, ferrara_2022_rebels_dustprops, sommovigo_2022_rebels_dusttemp, dayal_2022_rebels_dustmodel, topping_2022_rebels_stellarmass}, the SPT-selected DSFG catalog \citep{vieira_2013_spt,marrone_2018_spt0311, reuter_2020_spt}, and a selection of other galaxies at similar epochs, including both dusty starbursts and Lyman-break galaxies: A1689-zD1 \citep{watson_2015_A1689, knudsen_2017_a1689, bakx_2021_A1689}, A2744\_YD4 \citep{laporte_2017}, B14-65666\citep{hashimoto_2019_b14}, COS-87259 \citep{endsley_2022_SB_AGN}, and HFLS3 \citep{riechers_2013_hfls3, cooray_2014_hfls3}. We have selected galaxies that have robust measurements for redshift, stellar mass $(\gtrapprox 10^9)$, SFR, and dust mass for the most uniform means of comparison for the simulated galaxies. 

\begin{figure*}
    \centering
    \includegraphics[width=\textwidth]{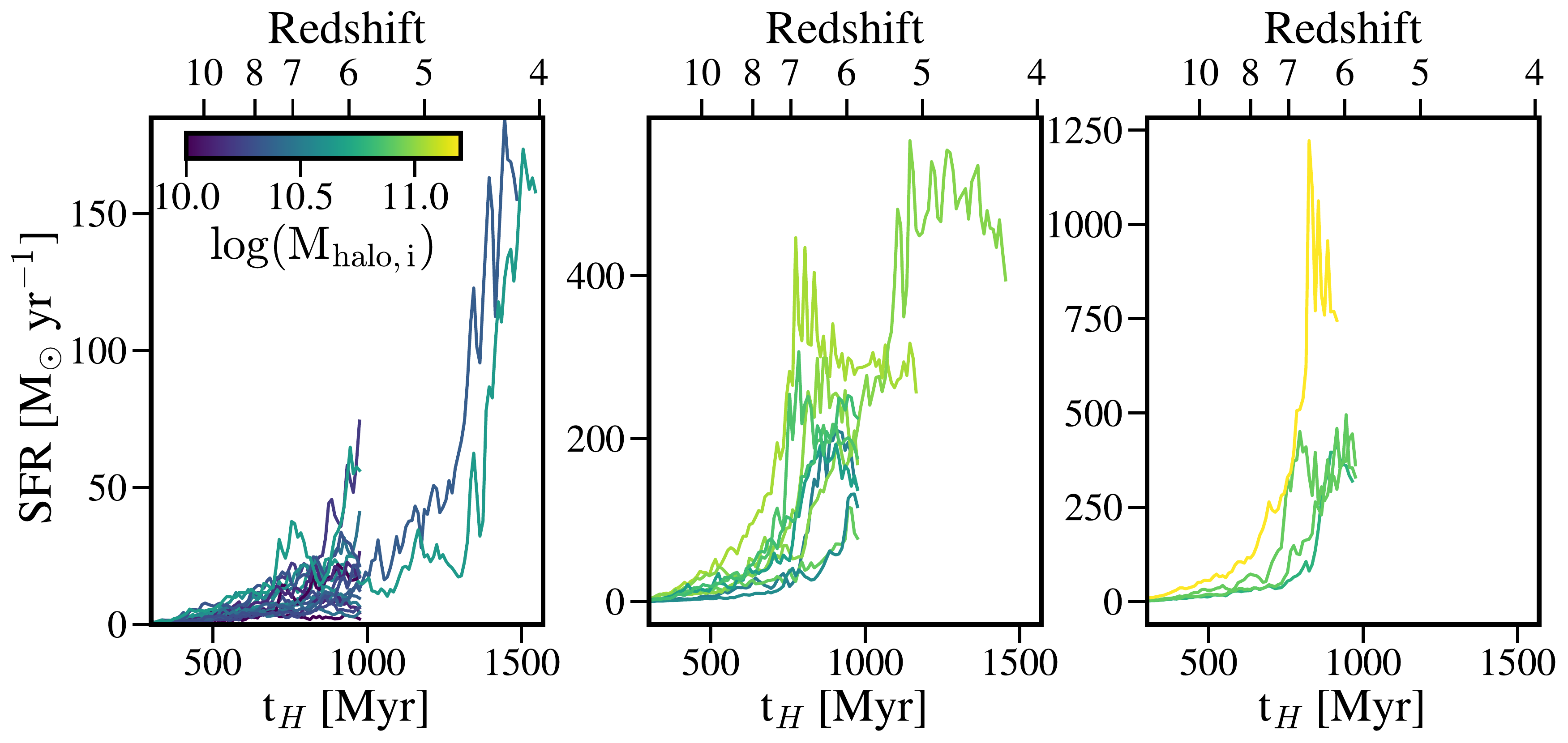}
    \caption{Cosmic Sands SFHs with galaxies binned according to their SFR at $z\sim6$ and colored by their initial halo gas masses. SFHs are diverse across halo mass but the maximum SFRs are found in the highest mass halos. The leftmost panel shows the relatively low mass galaxies that have SFRs $< 80 \: \mathrm{M}_{\odot}$ / yr at z=6. Two galaxies in this bin that have evolved further show that the SFHs continue to rise. The middle panel shows galaxies with $< 80$ SFRs $< 300 \: \mathrm{M}_{\odot}$ / yr and the right most panel shows the four most star forming galaxies.}
    \label{fig:sfhs}
\end{figure*}

\looseness=-1
In Figure \ref{fig:sfr_mass}, we plot the star formation rates and stellar masses of the simulated galaxies and the selected galaxies from the literature. For our zoom-in simulations, the infrared luminosities are calculated from the {\sc powderday} mock SEDs, integrated from $8-1000 \: \mu\mathrm{m}$. The physical properties are measured from the gas and star particles within the {\sc powderday} aperture, to ensure that the masses, SFRs, and luminosities are self-consistent with one another. This is in contrast to Figure \ref{fig:mass_dist}, in which the masses are calculated from only the particles that are bound to the central galaxy as determined from the {\sc caesar} friends-of-friends algorithm. Thus, there may be some minor but intentional inconsistencies between Figures \ref{fig:mass_dist} and \ref{fig:sfr_mass} which do not impact our analysis. The zoom-in galaxies span a wide range of stellar masses and star formation rates averaged over 50 Myr, ranging from $\log(M_*/M_{\odot})=3.7\times10^8$ to $\log(M_*/M_{\odot})=7.4\times10^{10}$ and SFR $=1.46 \: M_{\odot} \: \mathrm{yr}^{-1}$ to SFR $=937 \: M_{\odot} \: \mathrm{yr}^{-1}$. These values encompass the range of high-z galaxies from the literature and generally lie along the star-forming main sequence relation for $z = 6.7$. 

\looseness=-1
We note that our sample of zoom-in simulations are not necessarily a representative sample of massive galaxies in the first billion years but rather an illustrative one; the sample provides opportunities to study the difference modes of dusty, star-forming galaxy evolution. For that reason, we do not compare number counts or population statistics to observations from the literature. In what follows, we present the physical properties of our massive galaxies as well as mock spectral energy distributions (SEDs) and imaging generated by our {\sc powderday} radiative transfer. 

\begin{figure*}
    \centering
    \includegraphics[width=0.98\textwidth]{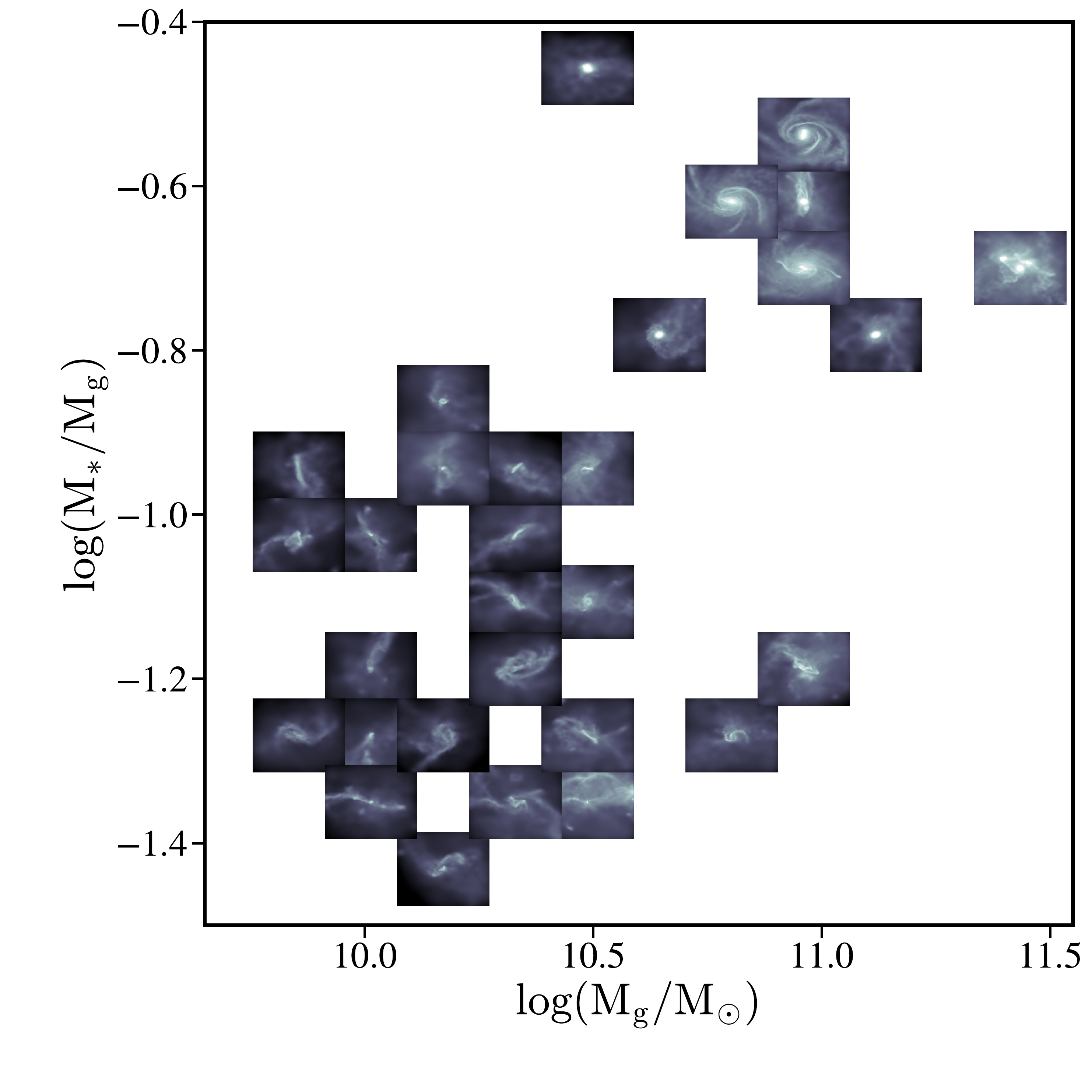}
    \caption{Early $(z = 6.7)$ Cosmic Sands morphologies as a function of their gas mass and stellar-to-gas mass ratio. The galaxies are represented by projections of their gas surface density; each thumbnail spans $3$ kpc with the observational angle chosen to align with the gas angular momentum vector. Galaxies with high gas masses have collapsed their gas into disks earlier, except for systems undergoing significant merger activity.}
    \label{fig:disks}
\end{figure*}

\subsection{Galaxy Physical Properties}\label{sec:properties}

\looseness=-1
In less than one billion years, nearly a third of our massive galaxy sample have reached a stellar mass of $10^{10}$ M$_{\odot}$ or larger. In this section we overview the star formation and dust mass evolution over cosmic time of our sample to understand the context of their position on Figure \ref{fig:sfr_mass}. In Figure \ref{fig:sfhs}, we show the SFHs for all galaxies, separated into bins according to their $z\sim6$ SFR and color-coded by their initial ($z=10$) parent halo masses. SFRs scale with halo mass and a majority of galaxy SFHs are rising over long timescales. While most of the simulations are run to $z=5.9$, some simulations have been evolved to lower redshift. The low-mass galaxies that have evolved to $z\sim4$ show steeply rising SFRs while higher mass galaxies appear to hit a plateau near $z\sim5$, similar to the trends seen in {\sc FLARES} galaxies \citep{wilkins_2022_flares_sfhs}. 

One major open question concerning early galaxy formation is whether or not massive early galaxies have significant star formation at times $z>10$ or if their stellar population is dominated by a single burst / sharply rising SFR near the time of observation. At $z=6$, we find that galaxies with sharply rising SFHs have mass-weighted stellar ages that are dominated by their recent star formation, i.e., these galaxies are younger compared to their less star-forming counterparts. However, all galaxy SFHs appear to be monotonically (on timescales $\gtrapprox 30$ Myr) rising, thus most of the galaxies' stellar populations are young ($< 100$ Myr) at these redshifts. Lastly, due to the resolution dependence and uncertainties in star formation and feedback modeling, specific predictions about the onset of star formation are difficult to make robustly.

\begin{figure*}[ht]
    \centering
    \includegraphics[width=0.98\textwidth]{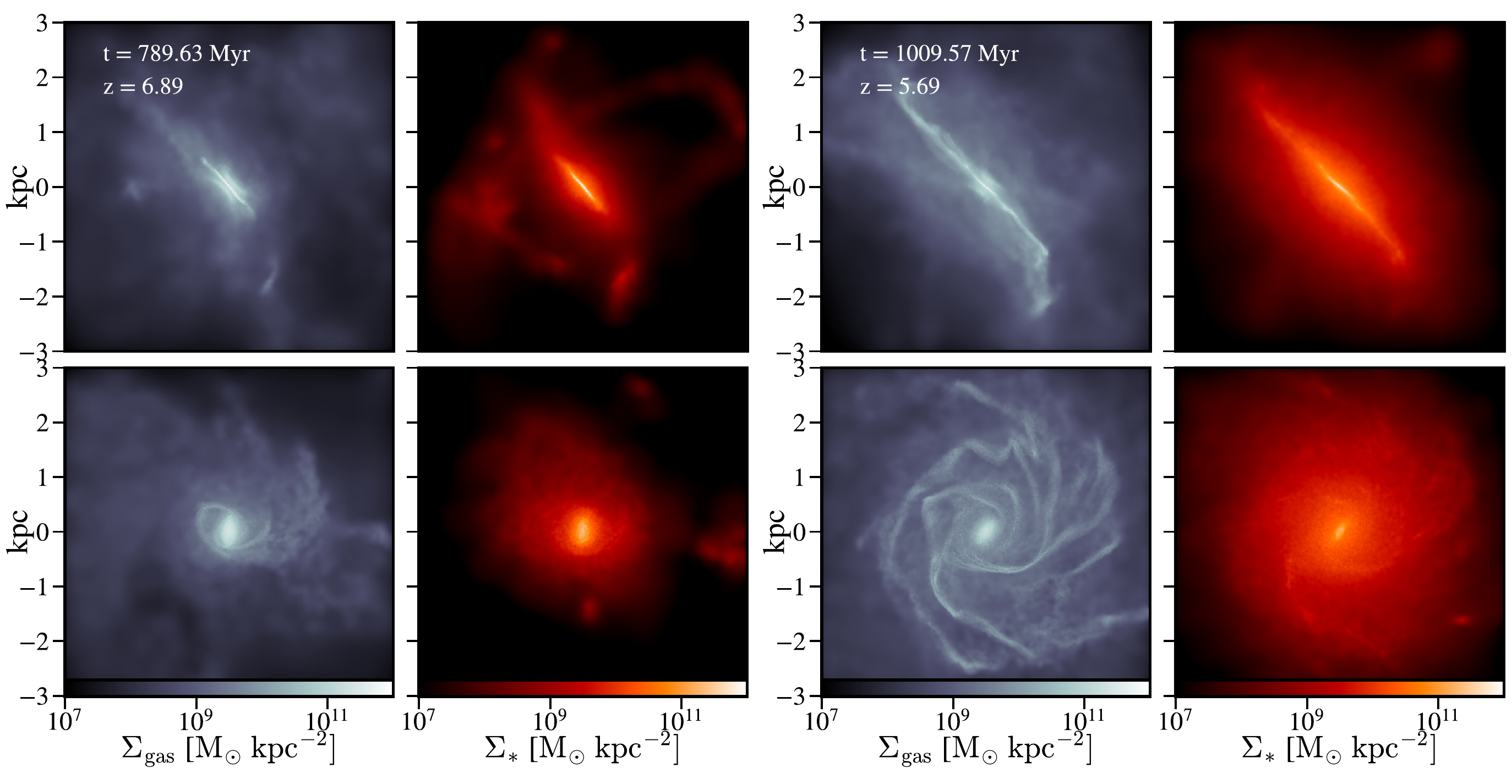}
    \caption{Example Cosmic Sands galaxy showing gaseous and stellar disk formation around $z=7$. Disk formation proceeds relatively quickly in the more massive Cosmic Sands halos. The rows show the galaxy at two viewing angles (edge on versus face on). The four panels on the left show the galaxy at $z=6.9$ and the four right panels show the galaxy $200$ Myr later at $z=5.7$.}
    \label{fig:disk_evolution_sigma}
\end{figure*}

Morphologically, the galaxies are compact with a significant fraction of galaxies settling into rotationally supported disks by $z=5$, with some systems forming a gas disk as early at $z=7$. In Figure \ref{fig:disks}, we show the range of early Cosmic Sands morphologies as a function of stellar-to-gas mass ratio and total gas mass; each galaxy is represented by a $3$ kpc thumbnail of the projected gas surface density at $z=6.7$. The most gas rich systems tend to have more ordered rotation earlier, except for systems that are undergoing significant merger activity. In \ref{fig:disk_evolution_sigma}, we show one such example of a compact disk galaxy formed at $z=6.9$. We plot the gas and stellar surface densities for two viewing angles (top row edge on, bottom row face on) and for two different times (left panel at $z=6.9$, right panel at $z=5.7$), showing the development of spiral arms over $200$ Myr. In Appendix \ref{appen:morph}, we show a gallery of galaxies spanning the stellar mass range of the Cosmic Sands sample. 

\begin{figure}
    \centering
    \includegraphics[width=0.48\textwidth]{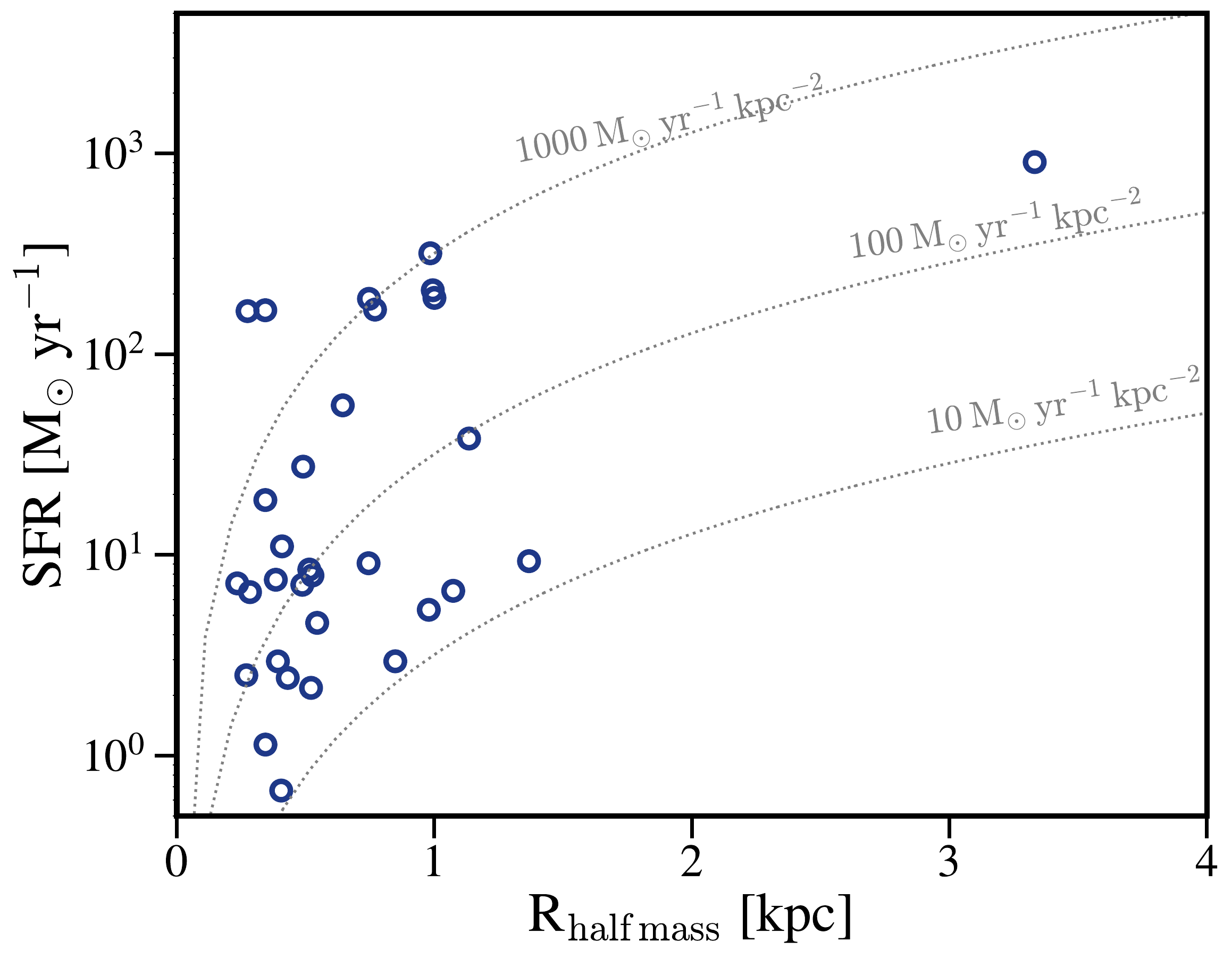}
    \caption{Star formation rate as a function of galaxy size for Cosmic Sands galaxies at $z=6.7$. The size is defined as the total half-mass radius as measured by {\sc caesar}, with the SFR equal to the instantaneous SFR within that radius. The gray dashed lines denote lines of constant SFR surface density. Most galaxies are compact with large SFR surface densities towards their nuclear regions.}
    \label{fig:sfr_surface_density}
\end{figure}

Star formation occurs primarily in the compact nuclear regions, resulting in the large stellar mass surface densities towards the center. We show the SFR surface densities in Figure \ref{fig:sfr_surface_density}, where we plot the SFRs as a function of galaxy total half-mass radius. The large SFR surface densities are comparable to observed values in high-z dusty, starburst systems \citep{ma_2016_spt_surface_densities,aravena_2016_spt_sfr, zavala_2018_dsfg_alma, zavala_2022_lensed_starburst_vla}.

\begin{figure}
    \centering
    \includegraphics[width=0.48\textwidth]{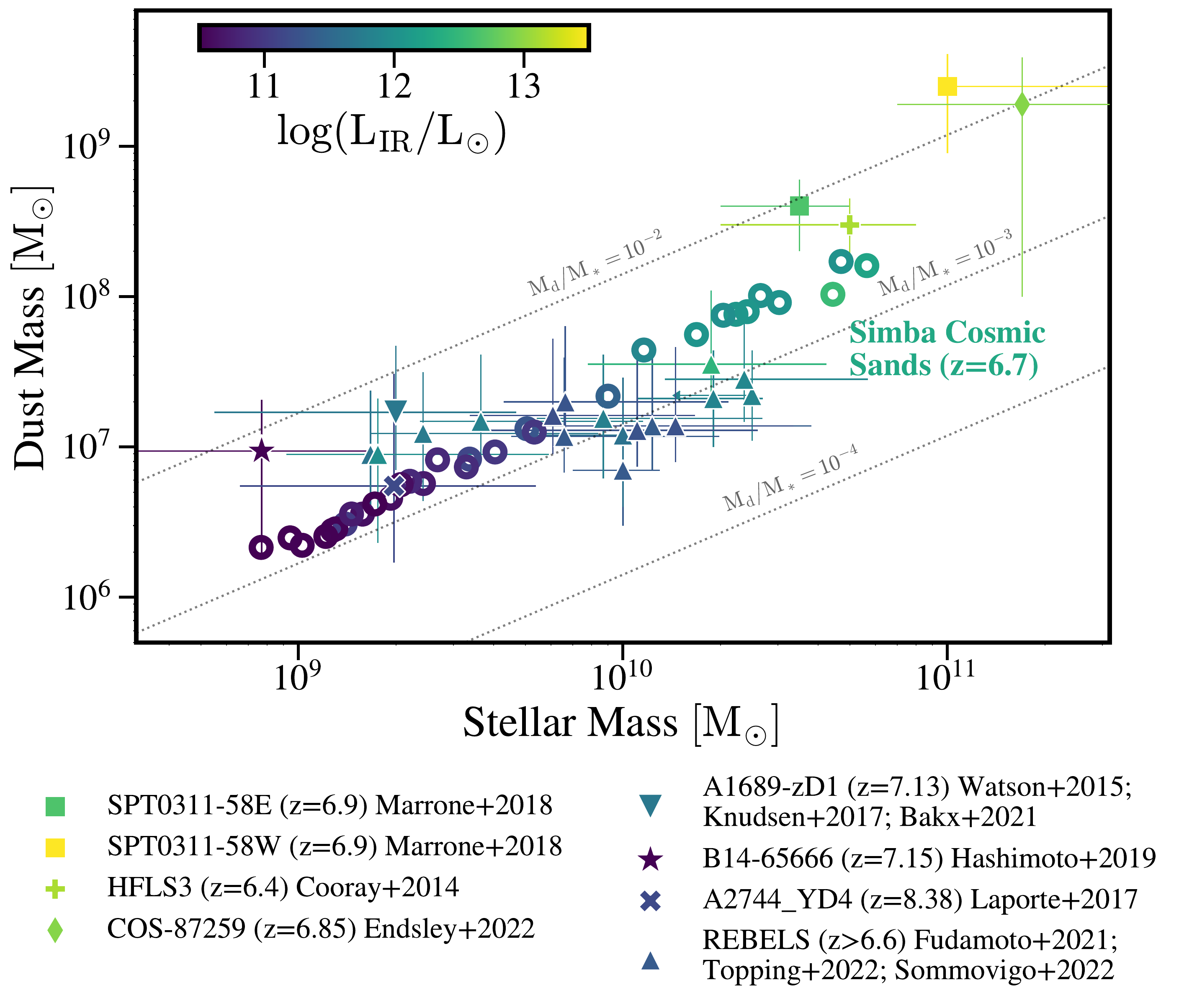}
    \caption{Dust mass as a function of stellar mass for the simulated galaxies and galaxies from the literature. Our Cosmic Sands galaxies span a dust-to-stellar mass ratio of $0.002 - 0.004$, consistent with the REBELs and other ``normal" dusty galaxies at this epoch. Symbols are colored by the total infrared luminosity. Dashed lines denote lines of constant dust-to-stellar mass ratios. Note that for galaxy COS-87259, L$_{IR}$ measurements exist for both an AGN component and thermal dust emission, where the latter value is plotted above.}
    \label{fig:dustmass_stellarmass}
\end{figure}

\looseness=-1
Resulting from these star formation rates, dust production primarily via supernovae (SNe) enables large reservoirs of dust to exist by redshift $6.7$. Hinted at in Figure \ref{fig:sfr_mass}, nearly a majority of the galaxies have infared luminosities  $\log(\mathrm{L}_\mathrm{IR} / \mathrm{L}_{\odot}) > 12$, comparable to ultra-luminous infrared galaxies (ULIRGs) found at lower redshift. These large infrared luminosities imply large swaths of dust are heated by the young stars produced in galaxies with SFRs $\gtrsim 100 \: \mathrm{M}_{\odot} / \mathrm{yr}$. In Figure \ref{fig:dustmass_stellarmass}, we show the dust masses as a function of stellar mass for the same galaxies in Figure \ref{fig:sfr_mass}. The Cosmic Sands galaxies exhibit a tight relation between their stellar and dust masses, with a dust-to-stellar mass ratio between 0.002 and 0.004, similar to the most massive systems in the original {\sc simba} $100$ Mpc/h box. The galaxies from the literature span a larger range of dust-to-stellar mass ratios, ranging from 0.0007 to 0.05. 

\subsection{Galaxy ``Observable" Properties}\label{sec:seds_imaging}

\begin{figure*}
    \centering
    \includegraphics[width=0.98\textwidth]{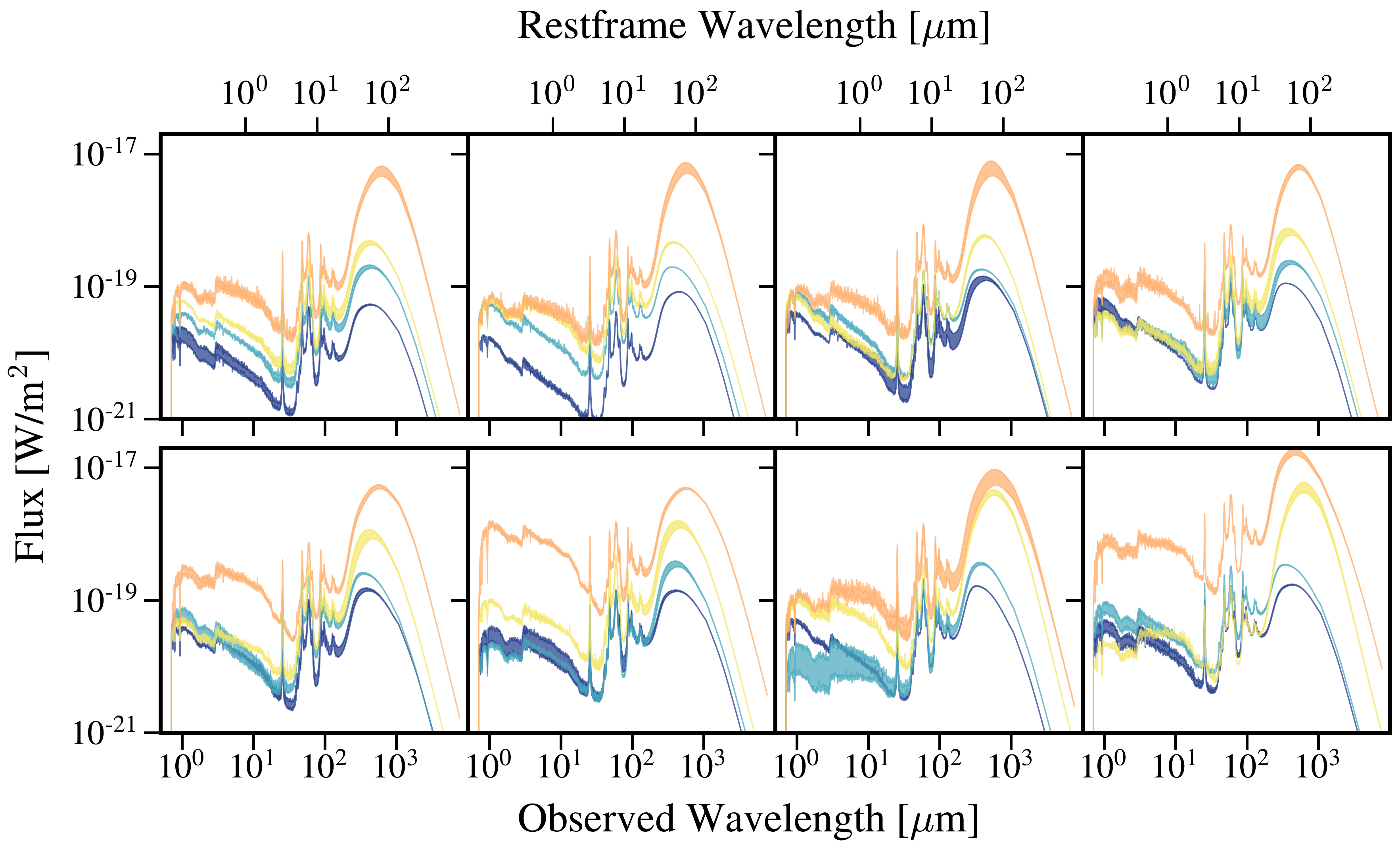}
    \caption{Spectral energy distributions (SEDs) for the central galaxies in each halo at redshift $z=6.7$. The panels separate galaxies by total luminosity with binning chosen to achieve maximum clarity between each SED. In each panel, galaxies are represented by different colors. We create the SEDs at 25 different viewing angles,  represented by the shaded regions for each galaxy. Some individual galaxies have a wide diversity in SED shapes as a function of viewing angle, thanks to inclination angle dependent dust attenuation.}
    \label{fig:seds}
\end{figure*}

\begin{figure*}
    \centering
    \includegraphics[width=0.32\textwidth]{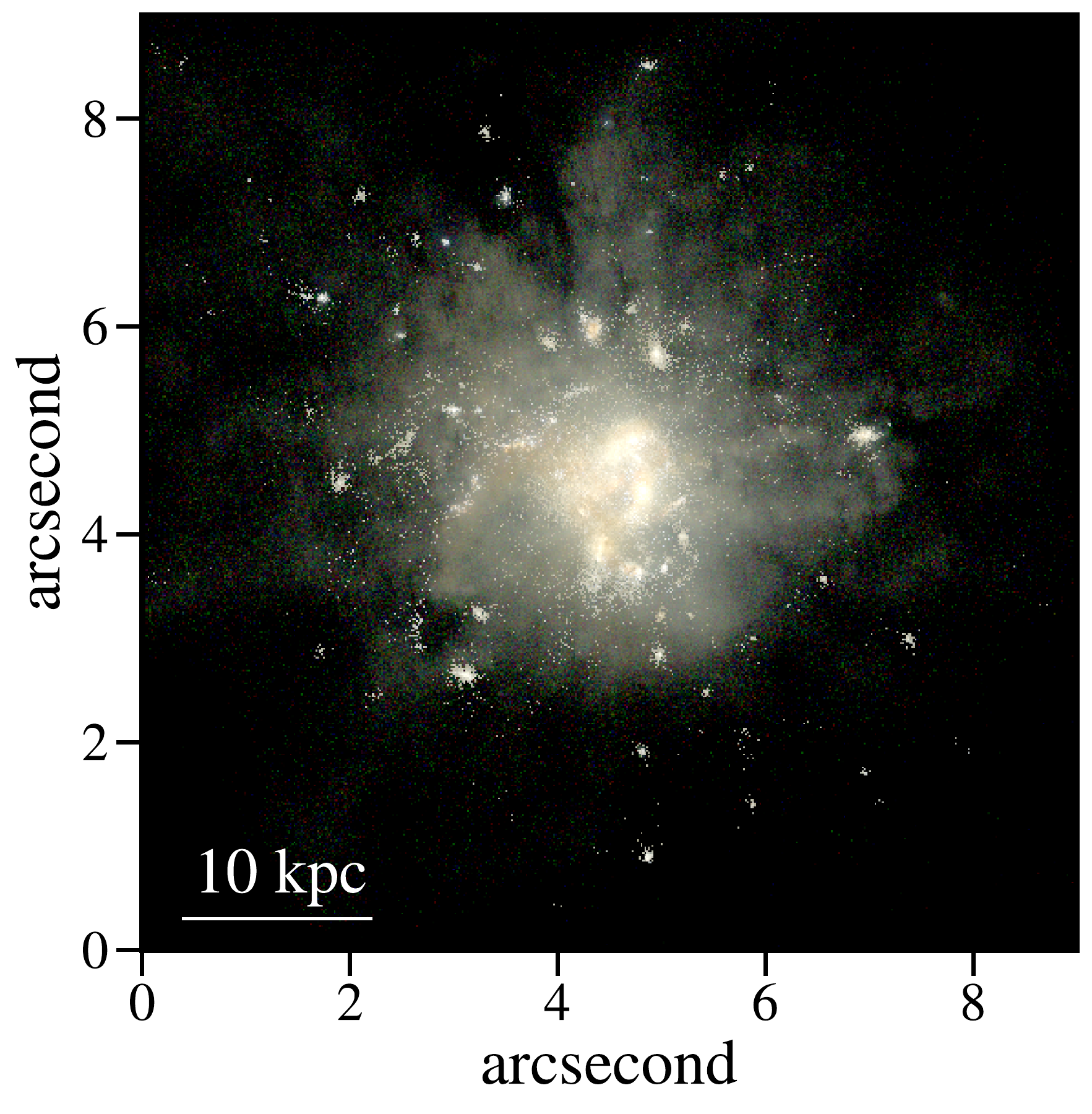}
    \includegraphics[width=0.32\textwidth]{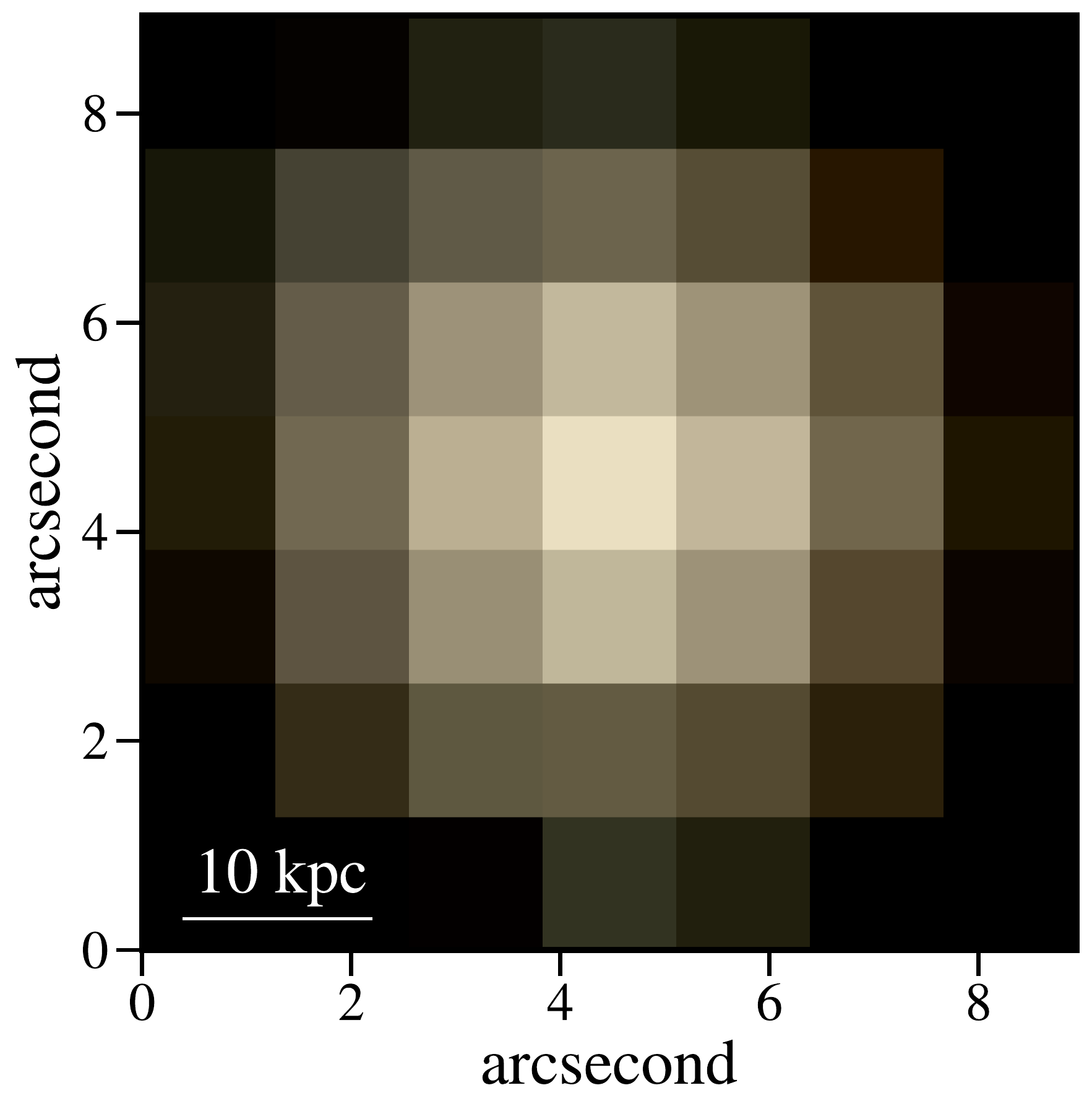}
    \includegraphics[width=0.32\textwidth]{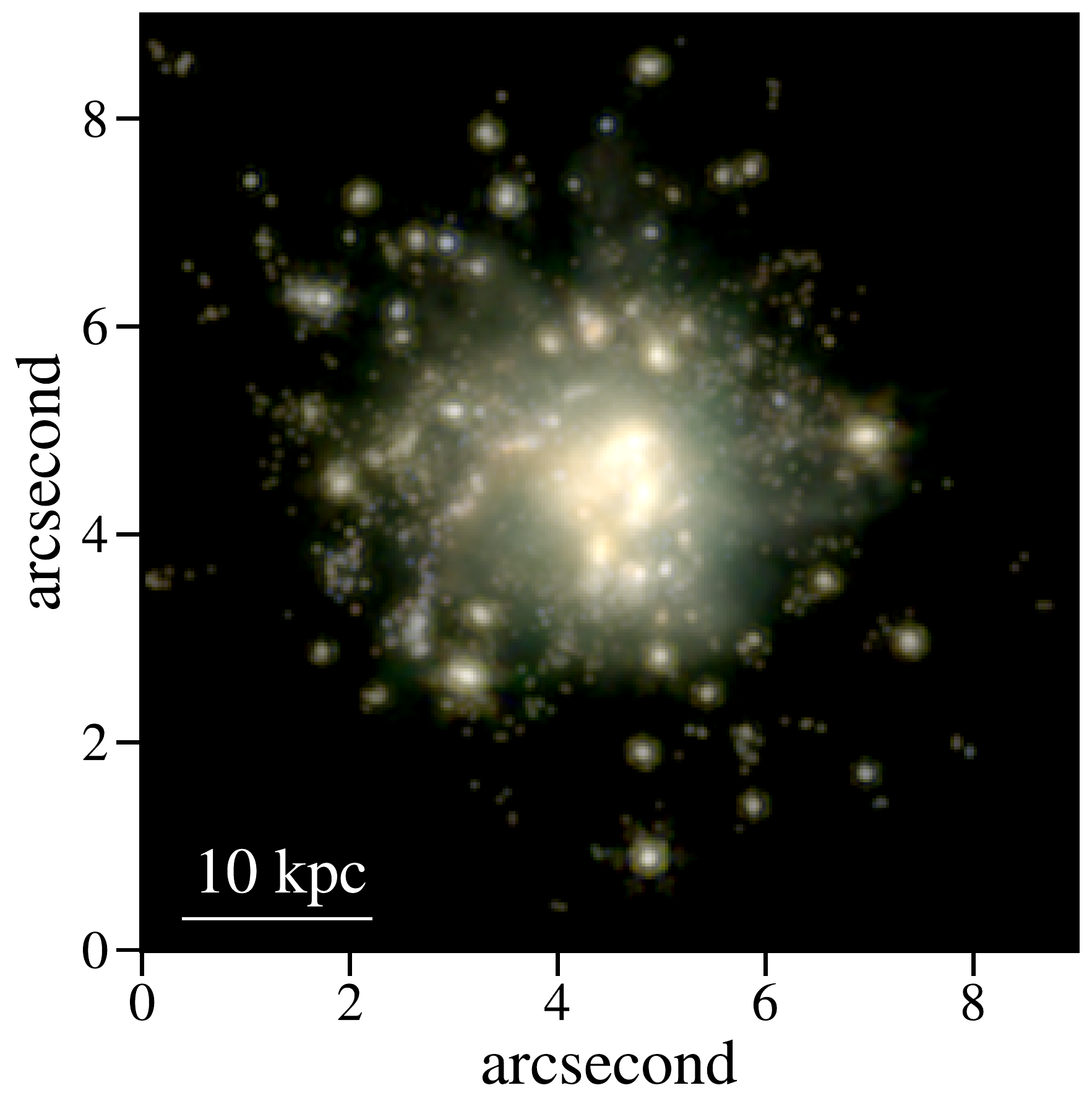}
    \caption{Mock RGB imaging of the merger-starburst system at redshift $z=6.62$, approximately $10$ Myr after in-fall of gas from the merger event. \textbf{Left}: Intrinsic image of the galaxy. \textbf{Middle}: Image convolved with \textit{Spitzer} IRAC PSF and pixel scale. \textbf{Right}: Image convolved with JWST NIRCam and MIRI PSF and pixel scales. The relatively poor angular resolution of the mock \textit{Spitzer} observation blends together the structure of the merging system while the mock JWST observations resolve the individual galaxies.}
    \label{fig:imaging}
\end{figure*}

\looseness=-1
We post-process our Cosmic Sands zoom-in simulations with 3D dust radiative transfer to infer the spectral energy distributions (SEDs) from the UV to the far-IR. This forward modeling allows us to represent our galaxies in the observable plane instead of the physical plane to allow for further comparisons to observations. In Figures \ref{fig:sfr_mass} and \ref{fig:dustmass_stellarmass}, we have shown the infrared luminosities calculated from these mock SEDs, for one viewing angle. In Figure \ref{fig:seds}, we show the galaxy SEDs binned by bolometric luminosity, such that galaxies are well separated in luminosity space for plot clarity. In each panel, the SED color denotes a galaxy, while the varying line thickness denotes SEDs generated from different viewing angles for each galaxy. For some galaxies, the SEDs are the same regardless of observer line-of-sight. For others, there is a spread of 0.1 - 0.2 dex in bolometric luminosity depending on viewing angle. This can be interpreted as an inclination dependent SED: for galaxies exhibiting disk-like morphologies, the observed SED will vary significantly if viewed from an edge-on sightline, where a large fraction of stellar light is obscured by dust,  versus the SED viewed from a face-on sightline, where the dust covering fraction is reduced. The morphology of a galaxy is in turn dependent on the stellar and dust growth history of the galaxy and parent halo, which we discuss in Section \ref{sec:gas_accretion}. Additionally, recent work by \citet{lovell_2021_orientation} has shown similar results (that the orientation of a galaxy influences its observed SED), resulting in potential orientation-angle driven selection biases. 

\looseness=-1
In addition to SEDs, we have performed monochromatic imaging of a subsample of our Cosmic Sands galaxies in several bands. In Figure \ref{fig:imaging}, we show a combined RGB image of the most massive system in our sample at redshift $z = 6.62$, corresponding to rest-frame wavelengths of $0.4,\:0.55, \: \mathrm{and}\:  0.8 \: \mu m$. The left-most image is the intrinsic image produced from the dust radiative transfer over a region of $50$ kpc. The middle panel shows this image convolved with the \textit{Spitzer} IRAC channels 1, 2, and 4 point spread functions (PSF), filter transmission curves, and pixel scales $(1.22 \: \mathrm{arcsec} / \mathrm{px})$, though we do not model instrument noise and assume perfect signal-to-noise. With relatively low angular resolution, the multi-component system is blended into an unresolved area of emission, from which determining the structure, kinematics, and physical properties of the galaxy would be challenging. 

\looseness=-1
The right-most panel of Figure \ref{fig:imaging} shows the intrinsic RGB image convolved with JWST NIRCam (277w, 444w) and MIRI (770w) transmission curves, PSFs, and pixel scales ($0.031$ and $0.1$ arcsec/pixel, respectively). With an angular resolution nearly 2 orders of magnitude greater than the IRAC image, the mock JWST image resolves the multi-component structure of the merger system.

\section{What Drives Intense Star Formation and Infared Luminosities in Early Massive Galaxies?}\label{sec:gas_accretion}

\looseness=-1
The Cosmic Sands simulation survey is intended to model massive galaxy evolution during the Epoch of Reionization. The physical properties of these modeled galaxies, as described in Section \ref{sec:properties}, include stellar masses, star formation rates, and dust masses comparable to the most extreme systems ever detected at high redshifts. These populations of galaxies imply intense growth and metal enrichment over timescales of a few hundred million years. In this section, we aim to understand the origin of their intense properties. Specifically, we ask the questions: What drives the extreme star formation in these massive systems? And do these SFRs power infrared luminosities comparable to observed galaxies? 

\begin{figure*}
    \centering
    \includegraphics[width=0.98\textwidth]{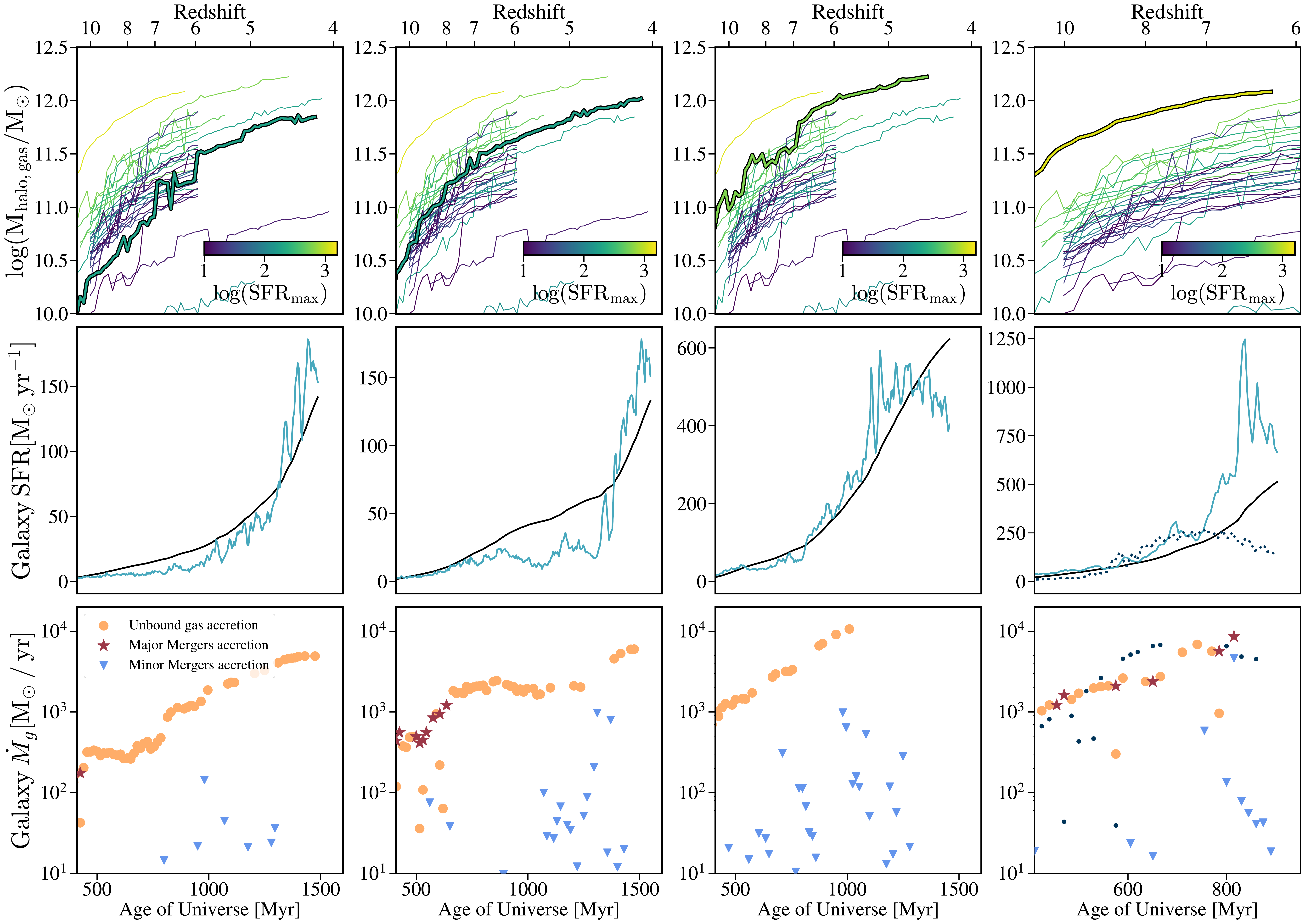}
    \caption{The growth history of selected Cosmic Sands galaxies. \textbf{Top}: Halo gas masses as a function of time for all halos in the sample. The four columns highlight different halos that span the gas mass range, from lowest halo mass on the left to the most massive halo on the right. \textbf{Middle}: The star formation history of the central galaxy associated with the parent halo highlighted in the top panel of each column. The black line shows SFR(t) compared to the main sequence in black, defined by the \cite{speagle_2014_sfms} slope and normalization set by the current stellar mass of the galaxy at each time. \textbf{Bottom}: Gas accretion histories for the selected central galaxies. The orange dots show the accretion of smooth gas (e.g., unbound),the maroon stars show major mergers, where the gas mass ratio of the interacting galaxies is $> 50\%$, and light blue triangles show the accretion from smaller minor mergers. The right most column shows the growth histories for the two nearly equal mass systems that occupy the halo. This is the only halo in our sample to host two such massive galaxies. The 'non-merger' system is shown by the dotted line in the middle panel and dark blue dots, showing the unbound gas accretion rate, in the bottom panel.}
    \label{fig:gas_accretion}
\end{figure*}

\subsection{Driving Early Star Formation}

Sustaining large SFRs requires a significant supply of gas. In high mass halos, two natural avenues for this gas accretion exist: filamentary accretion from the intergalactic medium \citep{Voort_2011_gas_accr, almeida_2014_accre}, as well as major and minor galaxy mergers. In our simulations, we track the gas accretion rates on the central galaxies as a function of mode -- accretion from filaments (and compact sub-halo structure) vs. accretion from major mergers. In practice, we define the contribution from major mergers by determining the bound progenitors of each galaxy over time and then computing the fraction of gas accreted from progenitors that are at least 50\% of the central galaxy's mass. Minor mergers are defined as the infall of {\sc caesar} identified subhalo structure that is less than 50\% of the central galaxy's mass onto the central galaxy.

In Figure \ref{fig:gas_accretion}, we show the gas accretion history for a small subsample of our galaxies, meant to represent the range of behavior seen across all galaxies in the simulation. We show for 4 galaxies the parent halo mass mass evolution, the galaxy SFH, and the galaxy gas accretion history. In the top row of Figure \ref{fig:gas_accretion}, the chosen parent halos (highlighted in bold) span the range of gas masses at redshift $z = 10$ and are color-coded by the maximum SFR achieved by the central galaxy. The halo gas mass distribution is the primary origin of the diversity of galaxy SFHs, shown in the middle row. The magnitude of SFRs increase from left to right, with the maximum starburst galaxy shown in the far-right column.

In each panel of the middle row, we plot the star formation histories of the selected galaxies (blue lines) alongside the main-sequence relation (black lines) as a function of time and current stellar mass, with the time dependent slope taken from \citet{speagle_2014_sfms} and the normalization equal to the cumulative stellar mass formed by that time. The star formation histories are calculated by binning the galaxy star particles by age and inferring the initial mass of each particle with {\sc fsps} stellar population modeling, given the current population age, metallicity, and mass. Both low- and high-mass galaxies tend to track the main sequence relation over time, with short ($<$10 Myr) starburst periods that temporarily push the galaxy over the median main sequence relation.  

In the bottom row of Figure \ref{fig:gas_accretion}, we show the galaxy gas accretion rates for the four selected galaxies. We split the accretion into two categories, depending on the source: 1) inflow of gas parcels not bound to any sub-halo structure (orange dots) 2) accretion of small, bound subhalos (light blue triangles) and 3) gas that is accreted onto the galaxy by way of major merger (maroon stars). The unbound gas accretion is the smooth gas accretion that tends to be steady with a rate set by the initial gas mass of the parent halo. Minor mergers are frequent for galaxies across the range of masses but their relative contribution to the overall gas inflow rate is typically small. The gas accreted via major mergers (where we define 'major' merger as an interaction between the central galaxy and a galaxy with at least 50\% of the gas mass of the central) is not necessarily tied to the gas mass of the parent halo, meaning that major mergers occur in halos regardless of mass. The impact of the major merger on the star formation rate of the central galaxy does appear to be a function of halo mass, as only galaxies in the most gas rich halos undergo gas rich major mergers with inflow rates comparable to the smooth accretion rates. 

The impact of a major merger on a galaxy's star formation is one that has been studied for many decades and the link between starbursts and mergers has largely been established \citep{toomre_toomre_1972,sanders_1988a_mergers, sanders_1988b_mergers,barnes_hernquist_1991_sbs_mergers, mihos_hernquist_1994_sb, mihos_hernquist_1996_mergers, barnes_hernquist_1996_mergers,sanders_mirabel_1996_ulirgs_araa, hopkins_2008_mergers, diMatteo_2007_mergers, cox_2008_mergers}. At high-redshift however, the relative importance of star formation fueled via major merger is less clear \citep{kartaltepe_2010_mergers_highz, Kaviraj_2013_highz_mergers, lofthouse_2017_highz_mergers, duncan_2019_highz_mergers, cibinel_2019_highz_mergers}; in principle, in early gas rich halos, large star formation rates could be fueled entirely by smooth gas accretion alone \citep{finlator_2006, dekel_2009}. Indeed, the compact nature and abundance of gas in early massive halos and galaxies could mean the conditions for starburst activity are met without merger induced torques driving gas inward. Furthermore, early major mergers may not have much impact on the proceeding star formation due to inefficiencies in the gas torquing in galaxies that have not yet established rotationally supported disks \citep{barnes_hernquist_1991_sbs_mergers, fensch_high-redshift_2016, renaud_2021_vintergatan_ii}.  

In our sample of model galaxies that do not undergo interactions between gas rich progenitors, the star formation rates achieved by the non-merger systems are substantially lower than their merger counterparts. The right-most column of Figure \ref{fig:gas_accretion} shows one such example of this: at redshifts ($z > 7$) the largest dark matter halo in our sample hosts 3 roughly equal mass galaxies. The merger of two of these galaxies around $820$ Myr sets off intense star formation which boosts the merger system well above the star-forming main sequence for a period of $20$ Myr. The merger progenitors were extremely compact proto-disk galaxies, which provided ample torque to funnel the gas into a nuclear starburst. The growth history of the third galaxy within the same halo that did not experience a major merger is shown in the dashed lines; while the smooth gas accretion rates onto the comparably compact disk and the resulting star formation history were comparable to the merger system prior to $800$ Myr, the galaxy-to-galaxy interaction made a significant difference between the two growth histories. We conclude that the most extreme star formation rates in early galaxies are the result of merger induced starbursts, but note that due to significant gas reservoirs, even secular SFRs are quite extreme with respect to the local Universe, as demonstrated in Figure \ref{fig:sfr_mass}. 

\begin{figure*}[t]
    \centering
    \includegraphics[width=\textwidth]{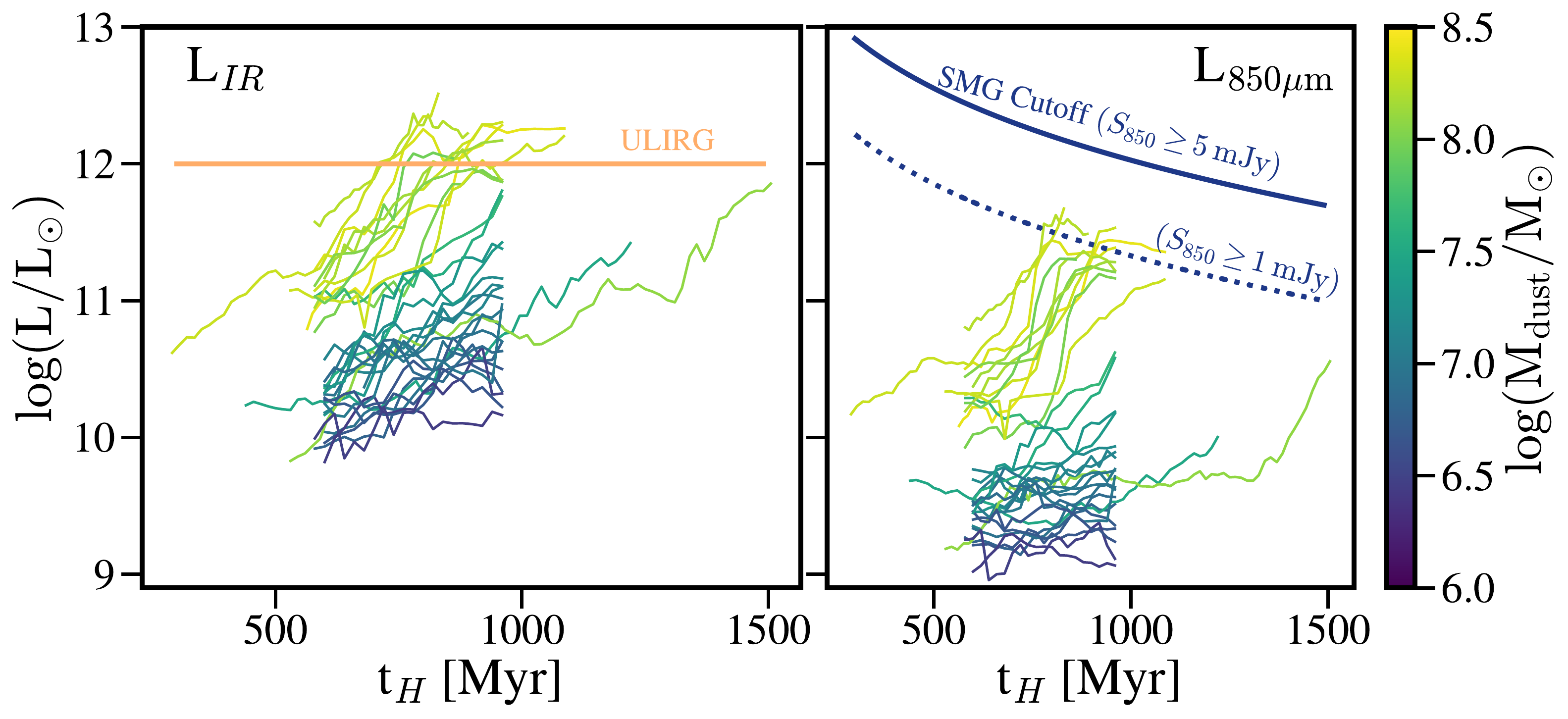}
    \caption{Luminosity evolution of the Cosmic Sands galaxies, showing that the most dust rich galaxies have ULIRG-like infrared luminosities over a prolonged period of time. \textbf{Left}: Total infrared $(8-1000 \: \mu\mathrm{m})$ luminosity as a function of time for the central galaxy in each halo. The lines are colored by the total galaxy dust mass at the latest snapshot. The orange line shows the ultra-luminous infrared galaxy (ULIRG) limit of $\log(\mathrm{L}_\mathrm{IR}) > 12$. \textbf{Right}: Observed-frame $850\mu$m luminosity as a function of time. The solid (dotted) blue line denotes the flux density cut-off for a galaxy to be classified as an SMG: $S_{850} > 5\: \mathrm{mJy}$ $(S_{850} > 1\: \mathrm{mJy}$).}
    \label{fig:smgs}
\end{figure*}

\subsection{Infrared Luminosities and Chemical Enrichment}

Though the local population of massive galaxies are largely dispersion dominated ellipticals that are seemingly devoid of star formation, the massive galaxy population of the early Universe was extremely dynamic. These galaxies include compact starburst systems, mergers, dust-obscured AGN, and sub-millimeter galaxies (SMGs). Even by redshift $z=6$, they are sufficiently chemically evolved to be observed in the rest-frame far-IR. But to what degree are these phases common to all massive galaxies? In other words, how common is it, in our Cosmic Sands sample, for a galaxy to be observable as an FIR-bright SMG? The answer depends on several factors, including the star formation history of the galaxy, as explored above, and the timescale of thermal dust emission.

\subsubsection{Defining an SMG}
 
The various observational methods through which dust-obscured galaxies are identified result in several ways of defining exactly what an SMG is. The first sub-millimeter bright galaxies were detected from the ground within narrow atmospheric transmission windows around $850 \: \mu\mathrm{m}$ and $850 \: \mathrm{mm}$. Those meeting a certain flux density cut-off (e.g., $\mathrm{S}_{850\mu\mathrm{m}} \geq  5 \: \mathrm{mJy}$ for sources targeted by the SCUBA instrument on JCMT \cite[e.g.][]{smail_1997_scuba, barger_1998_smgs, hughes_1998_smgs}) were classified as sub-millimeter galaxies \citep[SMGs, see review by][]{casey_narayanan_dsfgs_review}. Since then, the increased sensitivity of detectors on facilities like SPT and ALMA, as well as spaced-based telescopes targeting wavelengths inaccessible from the ground, have generally enabled classification of SMGs at fainter observed fluxes and further distances. One major limitation of SMG detections are the large beam sizes and low angular resolution of far-IR instruments, which results in source confusing and blending. For instance, the beam size $(0.5 \times \mathrm{FWHM})$ of the 250$\mu$m band of \textit{Herschel} SPIRE is $8.8"$, which corresponds to a spatial scale of approximately $76$ kpc at $z=2$, causing structures on smaller scales to become blended. With smaller beam sizes and higher resolution, SMGs can be de-blended to reveal that the most extreme systems are comprised of smaller interacting galaxies and structures \citep{hodge_2013_multiplicity, karim_2013_counts,hayward_2013_smg_blending, hayward_2018_multiplicity}. This is demonstrated in \S \ref{sec:seds_imaging} with the mock rest-frame optical and FIR imaging as well as in recent ALMA observations of a high-redshift SMG presented in \citet{spilker_2022_0311_alma}. In the context of comparisons between galaxy formation models and observations from the literature, consideration of the limitations of far-IR imaging is necessary to avoid biases in quantities like observed luminosity and sizes and to understand impacts on galaxy SMG classifications \citep{hayward_2011_smg_selection}.

Finally, the observed populations of SMGs are diverse; the extreme infrared luminosities observed in these systems can be the result of dust heated from both AGN and stellar sources. As noted in \cite{puglisi_2017}, SMGs can either be main sequence galaxies with more 'normal' global SFRs, in which case the large FIR luminosities could originate from a dust-obscured AGN, or they can be starburst galaxies, in which case the dust is heated by intense radiation from young stars. We choose to focus specifically on the impact that the star formation has on the dust thermal emission and determine if infrared bright phases are common amongst massive star-forming galaxies at $z>6$.

\subsubsection{Are Cosmic Sands Galaxies SMGs?}

Thermal emission from dust traces star formation over a longer timescale than other indicators (e.g., FUV luminosity and H$\alpha$ emission) and decreases at a slower rate compared to the drop off in star formation proceeding a burst  \citep{hayward_2014_fir_sfr, flore_velazquez_2021_timescales, ciesla_2021_fir_delay}. Therefore, dust heated by a starburst could keep infrared luminosities boosted to SMG-like values for longer periods, depending on dust surface densities and star-dust geometries. Thanks to the chemical enrichment from evolved massive stars, we have shown in Figure \ref{fig:dustmass_stellarmass} that the Cosmic Sands galaxies have dust masses comparable to both dust-obscured starburst systems, like SPT 0311-58E, as well as more `normal' $z=6$ star-forming galaxies, like A1689-zD1.

 We use {\sc powderday} to generate the infrared luminosities, assuming Milky Way-like dust optical properties (see \S\ref{sec:pd}) for all available snapshots. In Figure \ref{fig:smgs}, we show the evolution of the infrared luminosities of the Cosmic Sands galaxies as a function of time. The most dust rich systems (9 out of 32 systems) achieve luminosities comparable to ultra-luminous infrared galaxies (ULIRGs) by $z=8$ and maintain this IR brightness for over $500$ Myr. The timescales over which the galaxies are infrared bright are therefore much longer than their SFH burst timescales, which are on the order of $10-100$ Myr. IR bright phases are much less transient than starburst phases and appear to be a natural part of galaxy evolution for massive galaxies as suggested by \cite{narayanan_2015_smgs_nature} and \cite{lovell_2021_simba_counts}.

If we now consider the  observed-frame $850\mu$m luminosity as a means to compare to classical SMG selection criteria, none of the Cosmic Sands galaxies are above the $5$ mJy cutoff by $z=5.8$. If we lower the selection criteria to $1$ mJy, five of the most dust rich systems are above the threshold for $> 100$ Myr. At $z=6.7$, observed-frame $850\mu$m corresponds to a rest-frame wavelength of $110\mu$m, roughly straddling the contribution of warm and cold dust to the SED. Since none of the Cosmic Sands galaxies are above the $5$ mJy cut off, this could imply that the galaxies have warmer dust temperatures compared to classical SMGs. Additionally, we find that there are no galaxies above this flux threshold at $z>6$ in the {\sc simba} 100 Mpc/h volume, which matches the integrated SMG number counts and the number densities at $z=2$  \citep{lovell_2021_simba_counts}, suggesting that the Cosmic Sands simulation volume may be too small to form these extremely rare systems. So while the Cosmic Sands galaxies do not have the extreme IR / $850\mu$m luminosities of classical SMGs, their physical properties (SFR, mass), are remarkably similar to the rest of the observed sample.

\section{Discussion}\label{sec:discussion}

The Cosmic Sands galaxies are illustrative representations of the evolutionary cycles of massive galaxies. Formed within gas rich halos, these galaxies experience periods of concentrated gas accretion that fuels sustained star formation and starburst episodes, enabling the build-up of massive dust reservoirs and intense far-infrared emission. As presented in the previous section, while the formation history of these galaxies is diverse, a two-fold pathway emerges: smooth gas accretion can maintain SFRs above $250 \: \mathrm{M}_{\odot}/\mathrm{yr}$ but to achieve SFRs that boost galaxies well above the main sequence, a larger perturbation like a gas-rich major merger, is necessary to funnel enough gas into the galactic nucleus to trigger a starburst episode. And while the ULIRG-like infrared luminosities achieved by the most massive Cosmic Sands galaxies are impressive at such early times, none of the galaxies have bright enough FIR emission to be characterized as a classical SMG. In the following sections, we address follow-up questions that arise from our conclusions, including the fundamental uncertainties of our galaxy formation and stellar evolution modeling as well as comparisons to other models and a discussion on the future work we will do with the Cosmic Sands data set.

\subsection{Comparisons to Other Models}

Our understanding of early massive galaxy formation has progressed immensely during the last decade thanks to the advent of wide and deep observational surveys and the unprecedented sensitivity of facilities like ALMA and JWST. On the theoretical side, several modeling frameworks have made predictions for the earliest galaxy populations, focusing on the onset of star formation, chemical enrichment, and the contributors to the re-ionization of the Universe. These predictions are now being tested with JWST probing the $z>10$ redshift frontier, challenging the various modeling choices and uncertainties present in state-of-the-art simulations. Here, we compare our results to those from similar models, focusing on analysis of hydrodynamical simulations of early galaxies. 

The FirstLight simulations \citep{ceverino_2017_first_light}, with a co-moving spatial resolution of $\sim100$ pc, offer predictions for both low- and high-mass galaxies as cosmic dawn. They find that the SFHs of galaxies across a range of stellar masses are diverse but most galaxies undergo several short ($\lessapprox 100$ Myr) bursts of star formation early on in their formation, similar to the Cosmic Sands galaxies, as the similar time resolution of the snapshots can resolve short timescale SFR variations \citep{ceverino_2018_first_light_ii}. Since the Cosmic Sands galaxies were chosen from the most massive halos, the galaxy property parameter space overlap between FirstLight and Cosmic Sands galaxies is limited, but the high mass end of the FirstLight sample has comparable SFRs and gas depletion timescales to those from Cosmic Sands.

The FLARES simulations \citep{lovell_2021_flares}, with a baryonic mass resolution of $\sim10^6$ M$_{\odot}$, model a range of over-densities in the epoch of re-ionization to explore the environmental dependence of early galaxy formation. The high mass end of the FLARES stellar mass function (M$_* > 10^9$) has comparable SFRs to the Cosmic Sands galaxies but due to our limited sample size, it is difficult to determine if the Cosmic Sands galaxies show the same high mass star-forming main sequence turnover. Analysis of SFHs in \cite{wilkins_2022_flares_sfhs} shows remarkable similarities between the FLARES and Cosmic Sands galaxies, namely that lower mass galaxies have primarily rising SFHs towards $z=5$ while higher mass galaxies appear to hit a plateau. 

Finally, {\sc vintergatan} is a high resolution zoom-in simulation of a Milk Way-like galaxy \citep{agertz_2021_vintergatan_i}. While on the lower mass end of the Cosmic Sands stellar mass range, at $z>4$, the Milky Way progenitor has a rising SFR and experiences a continuous stream of gas accretion from bombardment of small subhalos and gas streamers. The galaxy's thick disk does not form until $z\sim2.7$, and so major mergers do not impact the SFR significantly \citep{renaud_2021_vintergatan_ii}. Similarly, the higher mass Cosmic Sands galaxies tend to form disks sooner while the lower mass galaxies remain unorganized and are less impacted by mergers.

\subsection{Modeling Uncertainties}\label{sec:caveats}

\begin{figure}[t]
    \centering
    \includegraphics[width=0.47\textwidth]{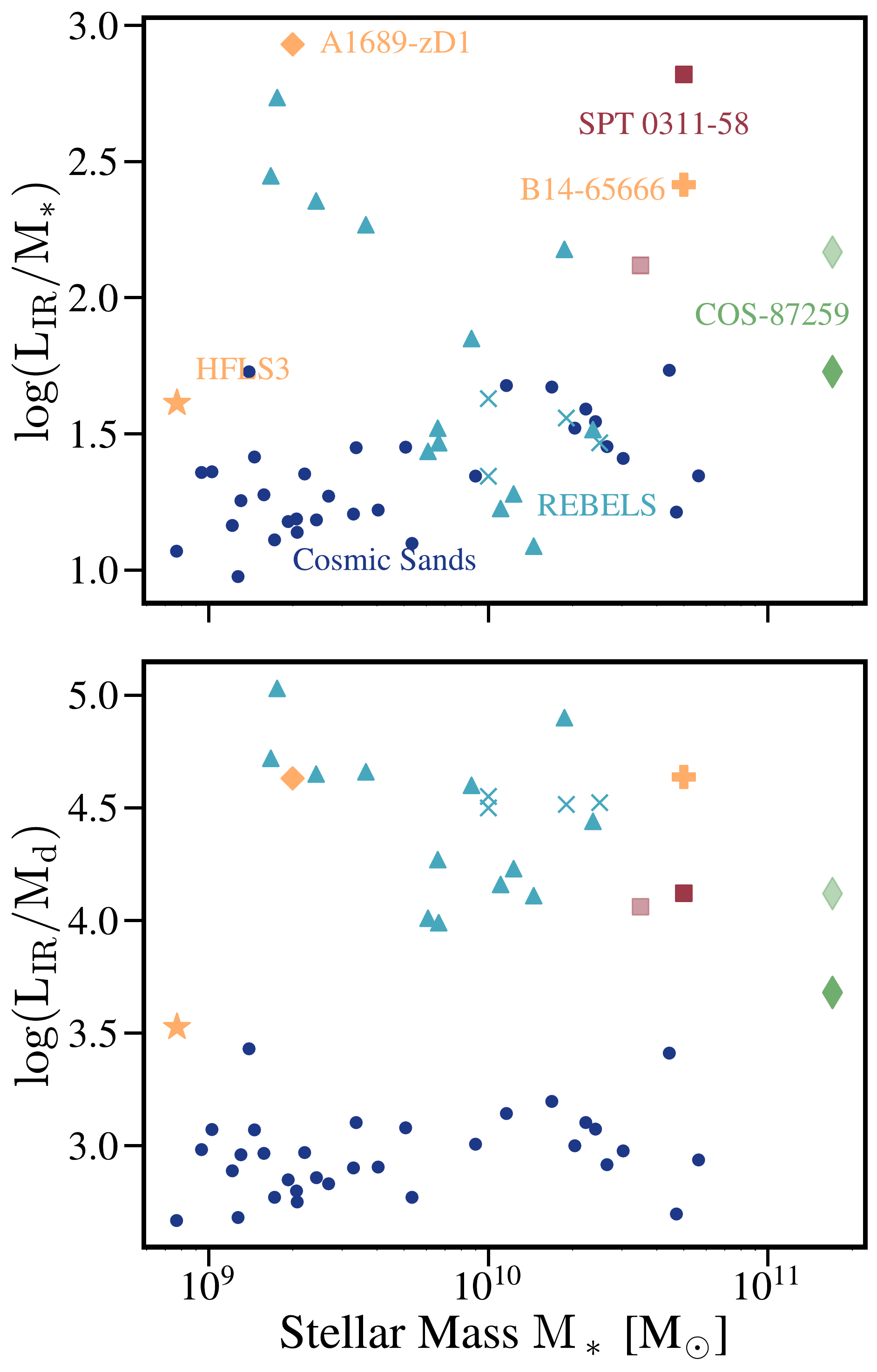}
    \caption{Ratio of galaxy infrared luminosity to stellar mass (top) and dust mass (bottom). The Cosmic Sands galaxies are shown in dark blue points, while the other points are the observations comparison sample from Figures \ref{fig:sfr_mass} and \ref{fig:dustmass_stellarmass}. The Cosmic Sands galaxies occupy a narrow range of luminosity-to-mass ratios while the observations are more diverse at fixed stellar mass. The western (eastern) source of SPT 0311-58 is shown in dark (light) red. COS-87259 \citep{endsley_2022_SB_AGN} has contributions to the IR luminosity from both dust (darker green point) and an obscured AGN (lighter green).}
    \label{fig:LIR_mass}
\end{figure}

\subsubsection{Dust Properties and Infrared Luminosities}\label{sec:dust_unc}

The success of the Cosmic Sands simulations to reproduce the population of extreme massive systems in the early Universe is primarily due to the galaxy formation physics within {\sc simba} as well as the ability to self-consistently model dust growth and star-dust geometries. As shown in Section \ref{sec:properties}, our galaxies have comparable stellar and dust masses and star formation rates to systems observed at similar epochs. However, the radiative properties of the galaxies, specifically the infrared luminosities, are in mild tension with the observations. For example, at a fixed stellar mass, the infrared luminosities from the simulated galaxies occupy a narrower range than those observed. We show this explicitly in Figure \ref{fig:LIR_mass}, where we plot the light-to-mass ratios for both stellar and dust mass as a function of stellar mass. The Cosmic Sands galaxies (dark blue) occupy a narrow range in light-to-mass ratio over the range of galaxy stellar mass. 

Though the differences are slight, the implications of the lack of overlap in parameter space are potentially significant. For instance, several uncertainties in the forward modeling of the Cosmic Sands galaxies exist, including the stellar population modeling (the IMF, stellar isochrones, and the stellar spectral library chosen to model the luminosity of the star particles, see, e.g., \cite{akins_2022_uvj}) as well as the assumed optical properties of the dust grains and thus the underlying extinction curve of the galaxy spectrum. Modifying the optical properties of the dust grains would propagate changes in the resulting infrared luminosity for the same dust and stellar mass values. Constraints on the optical properties of dust at high-redshift are not widely available, thus we elect to use a fixed dust extinction curve that matches the average Milky Way curve. However recent numerical studies including an evolving dust grain size distribution in \citep{li_2020_mw_ext_curve, makiya_hirashita_2022_grain_size_evo} have shown that galaxies can initially exhibit steep dust extinction curves that then evolve toward shallower UV-optical slopes as the grain size distribution evolves to match the MRN \citep{mrn} distribution. 

At the same time, there may be significant uncertainties in the observed infrared luminosities of high-$z$ galaxies.  For example, if the FIR SED is not fully sampled -- a scenario that is common -- then the derived infrared luminosity is steeply dependent on the highly-uncertain dust temperature.  Only with a fully sampled UV-IR SED can the dust properties be properly inferred, given the degeneracies between temperature, mass, and absorption properties. 

Furthermore, though {\sc simba} has been shown the match the dust properties of $z = 0-2$ galaxies \citep{qi_dust, dud_2021}, uncertainties remain with regards to modeling the build up of dust in the early Universe, including the processes of dust production from stellar sources and dust growth in the ISM. While systematically testing various dust models is beyond the scope of this work, in a future study we will focus on analyzing these uncertainties and degeneracies in dust modeling with a particular focus on the efficiency of dust growth in the Epoch of Reionization.

Lastly, we note that we do not model any active galactic nuclei (AGN) emission for the Cosmic Sands galaxies. Thus, the measured infrared luminosities are a result of reprocessed starlight alone. Recently, however, \cite{endsley_2022_SB_AGN} reported luminosity estimates for COS-87259 with contributions from an AGN and dust. These two values are shown in Figure \ref{fig:LIR_mass}. The dust-only infrared measurement (dark green) is more in agreement with the light-to-mass ratios of the Cosmic Sands galaxies. This could imply that some of the discrepancies in our luminosity values originate from neglecting AGN emission. Future work will be done to analyze the importance of AGN in early, massive galaxy growth.

\subsubsection{Stellar Feedback}
While the {\sc simba} suite of galaxy formation physics allows us to self-consistently model complex galaxy growth histories and dust mass build up in massive halos, as a necessity, processes like star formation and the internal physics of molecular clouds are not explicitly modeled due to resolution constraints. For instance, star formation and the structure of the ISM are modeled with an effective equation of state in that is tuned to reproduce the \cite{kennicutt_1989_kslaw} - \cite{schmidt_1959} relation. The star formation histories that result from an effective equation of state have been shown to be steadier and less prone to burstiness compared to models, like {\sc fire} \citep{hopkins_FIRE_og} and {\sc smuggle} \citep{marinacci_2019_smuggle}. 

As shown in \cite{hopkins_FIRE_og}, the explicit modeling of stellar feedback processes results in SFHs that are quantitatively different than the SFHs produced by models employing sub-grid resolution prescriptions, owing primarily to the coupling of the various feedback processes that is not reproducible by only characterizing the net effects of feedback. Namely, the SFHs produced by the FIRE model peak in star formation later and have greater short timescale variability compared to models without explicit feedback models \citep{iyer_2020}.

This represents a fundamental uncertainty in modeling massive galaxy formation and evolution; the 'burstiness' of real galaxy star formation histories is itself an uncertain quantity due to the challenges of inferring small timescale variations in the SFH and even large starburst episodes from spectral energy distribution modeling \citep[e.g.][]{iyer_2019_bursts}.

\section{Summary and Conclusions}

In this paper, we present the Cosmic Sands sample of massive galaxies in the Epoch of Reionization. Built on the {\sc simba} model for galaxy formation, we produce 32 dark matter halos with gas rich central galaxies featuring SFRs as high as $1000 \: \mathrm{M}_{\odot} \: /\mathrm{yr}$. We show that the Cosmic Sands galaxies span a wide range of stellar masses and star formation rates, matching the ``normal" star-forming dusty galaxy population of, e.g., REBELS \citep{bouwens_rebels, topping_2022_rebels_stellarmass} as well as the extreme starburst and SMG systems of HFLS3 \citep{riechers_2013_hfls3} and SPT 0311-58 \citep{marrone_2018_spt0311}. Below we summarize the notable findings of our analysis:

\begin{enumerate}

    \item At redshift $z=10$, proto-massive galaxies are extremely compact and undergo nearly continuous bombardment of gas accretion from both smooth gas streams and mergers with other subhalos. But by $z=6$, some systems have evolved into disk-like galaxies with distinct gas dense spiral arms and compact star formation in their nuclei.
    \item Smooth gas accretion can maintain SFRs above $250$ M$_{\odot}/$yr but to achieve SFRs that boost galaxies well above the main sequence, a larger perturbation like a gas-rich major merger is necessary to trigger a starburst episode. Thus we conclude from \S \ref{sec:gas_accretion} and Figure \ref{fig:gas_accretion}, that, while rare, dusty starburst galaxies in the early Universe are primarily merger driven. 
    \item Coupling the Cosmic Sands simulations with dust radiative transfer, we find that the infrared luminosities at $z\sim6$ of the most dust rich systems are comparable to local ULIRGs but are substantially dimmer than the extreme SMG systems. Since the physical properties of the galaxies are comparable to the observed sample of galaxies, we suggest that this discrepancy could be due to modeling uncertainties (dust optical properties, stellar feedback models) and difficulties in measuring dust properties of observed galaxies, as discussed in \S \ref{sec:dust_unc}.
\end{enumerate}

The galaxies in the Cosmic Sands sample are an ideal laboratory to study in detail the processes that influence massive galaxy formation and evolution. While not wholly representative of all early massive galaxies, they are rather illustrative examples highlighting the various pathways galaxies take during their evolution, from major merger events to starbursts and SMG phases. The primary question we aim to address with Cosmic Sands is under what conditions do massive, dusty, and extreme systems form and what is their subsequent fate? 

One major question concerning early massive, dusty galaxies is how do they achieve such large dust-to-stellar mass ratios. Studies into the predicted dust yields from Type II SNe and winds from evolved stars reveal that theoretical estimates are in tension with observed dust masses in galaxies at $z>6$ \citep{michalowski_2015_dust, ginolfi_2018_dust,lesniewska_2019_dust_yields}. Lack of constraints on basic dust properties such as composition and grain size distribution, due to uncertainties about the ISM conditions that drive the growth and destruction of dust grains, make modeling the build up of dust challenging \citep{burgarella_2020_early_dust}. 

While the {\sc simba} galaxy formation framework has been shown to reproduce dust properties at redshifts $z=0-2$, what remains unclear is the contribution of the various dust formation pathways to the overall dust content in a galaxy, especially at higher redshifts. We aim to address these uncertainties in future studies by comparing various methods for dust implementations in cosmological simulations.

\acknowledgements
The authors would like to thank Sune Toft, John Weaver, and Gergo Popping for fruitful discussions that guided the analysis of this paper, and Dan Stark and Michael Topping for sharing their stellar mass measurements for the REBELs sources. The Cosmic Dawn Center is funded by the Danish National Research Foundation under grant No. 140. SL and DN acknowledge support from the NSF via grant AST-1909153.

\appendix

\section{Morphologies of $\lowercase{z} \sim 6$ Galaxies}\label{appen:morph}

Here we present gas- and stellar-mass surface densities at edge on and face on observing angles for a subsample of Cosmic Sands galaxies to compliment the galaxy shown in Figure \ref{fig:disk_evolution_sigma}. In each figure below, the four left hand panels show the surface densities at an early epoch and the right hand panels show the densities at a later epoch for the same galaxy. Observing angles are chosen based on the angular momentum vector of the gas in the galaxy as a proxy for the normal vector, though in some systems without ordered rotation, the normal vector is not well defined and ``edge on" and ``face on" views are chosen by eye.

\begin{figure*}
    \centering
    \includegraphics[width=0.48\textwidth]{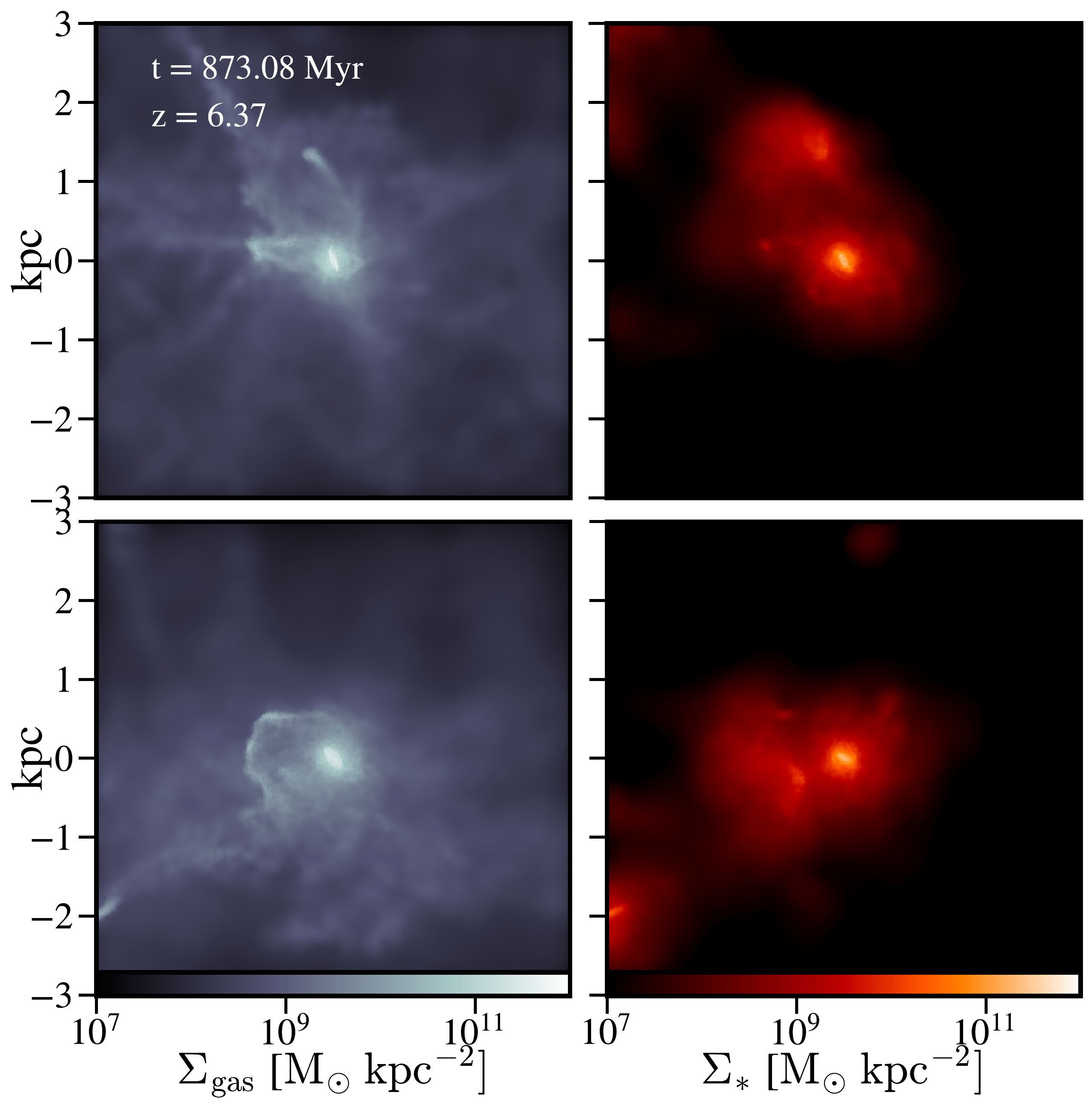}
    \includegraphics[width=0.48\textwidth]{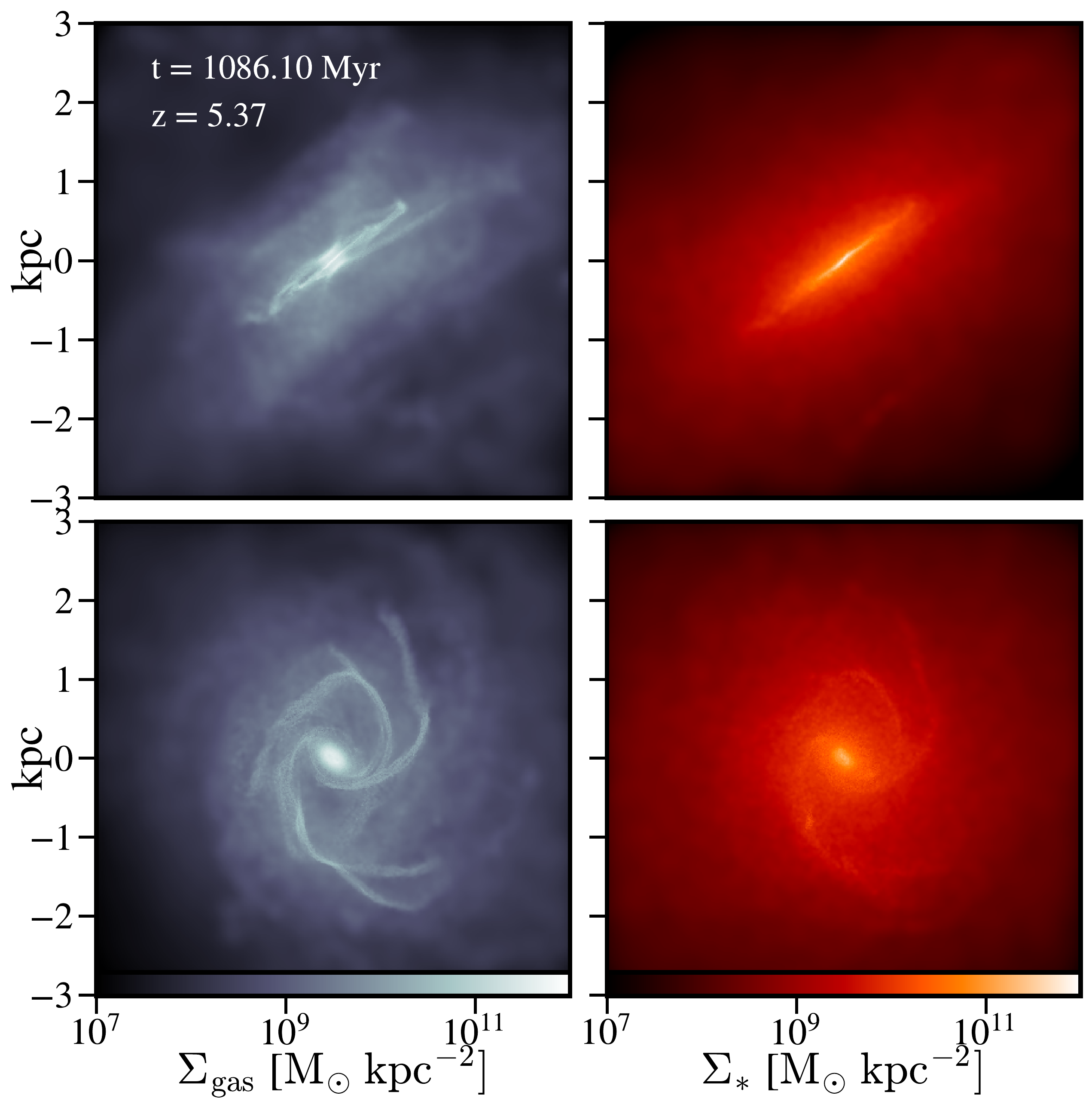}
    \caption{Same as Figure \ref{fig:disk_evolution_sigma} but for a different galaxy. The rows show the galaxy at two viewing angles (edge on versus face on). The four panels on the left show the galaxy at $z\sim6.4$ and the four right panels show the galaxy $200$ Myr later at $z\sim5.4$.}
    \label{fig:run0_morphology}
\end{figure*}

\begin{figure*}
    \centering
    \includegraphics[width=0.48\textwidth]{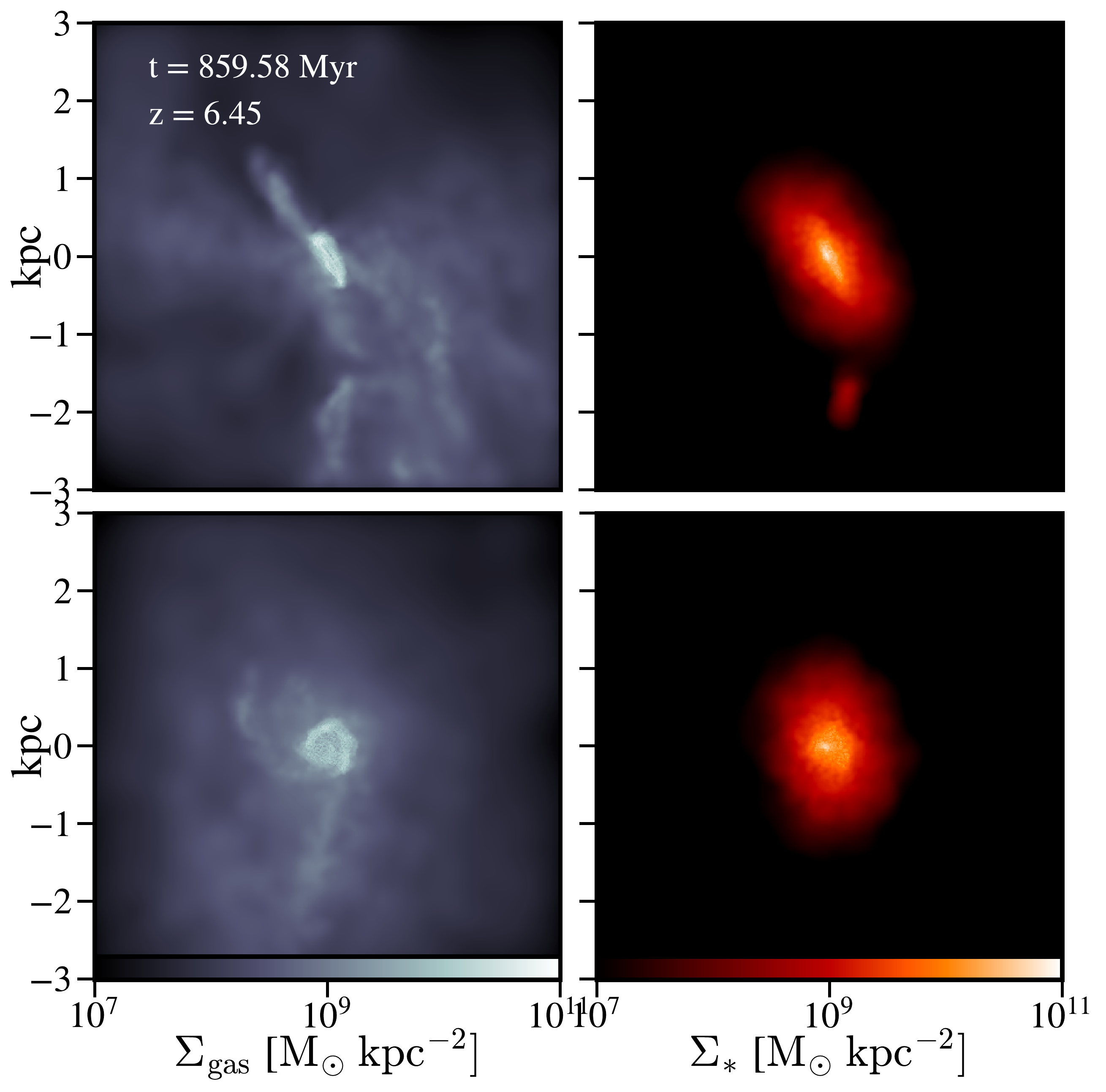}
    \includegraphics[width=0.48\textwidth]{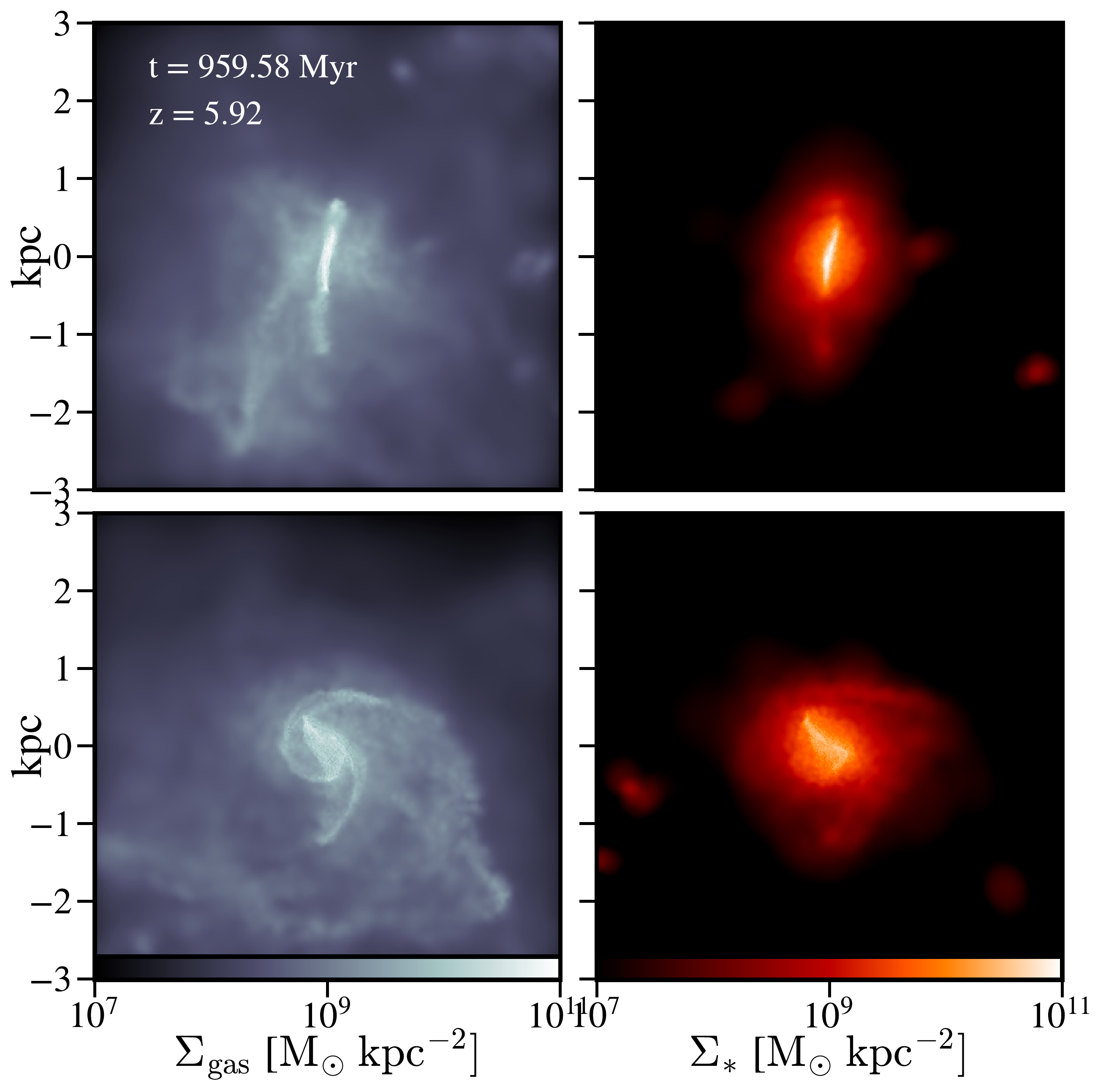}
    \caption{Same as Figure \ref{fig:run0_morphology} but for a different galaxy.}
    \label{fig:run3_morphology}
\end{figure*}

\begin{figure*}
    \centering
    \includegraphics[width=0.48\textwidth]{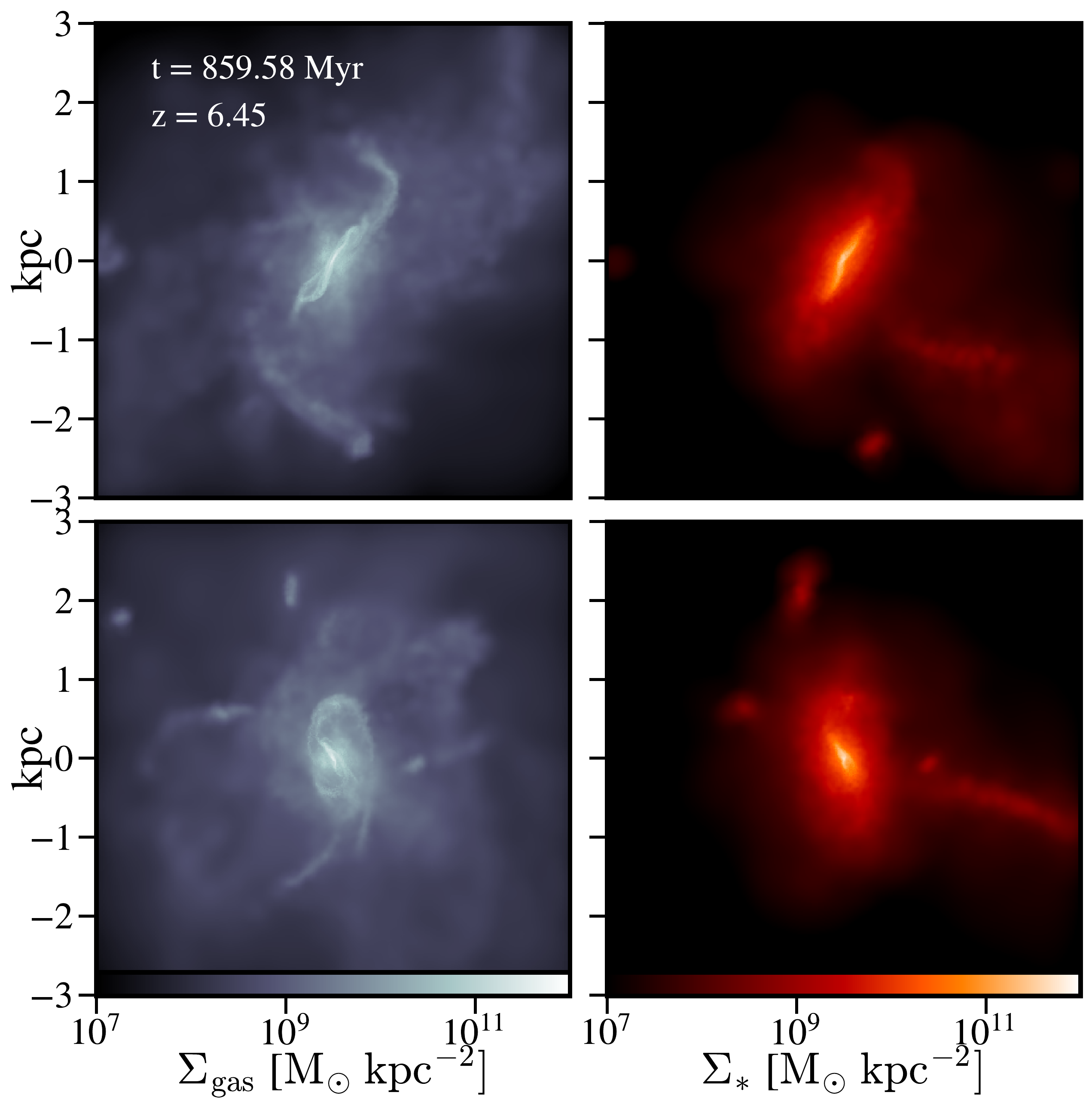}
    \includegraphics[width=0.48\textwidth]{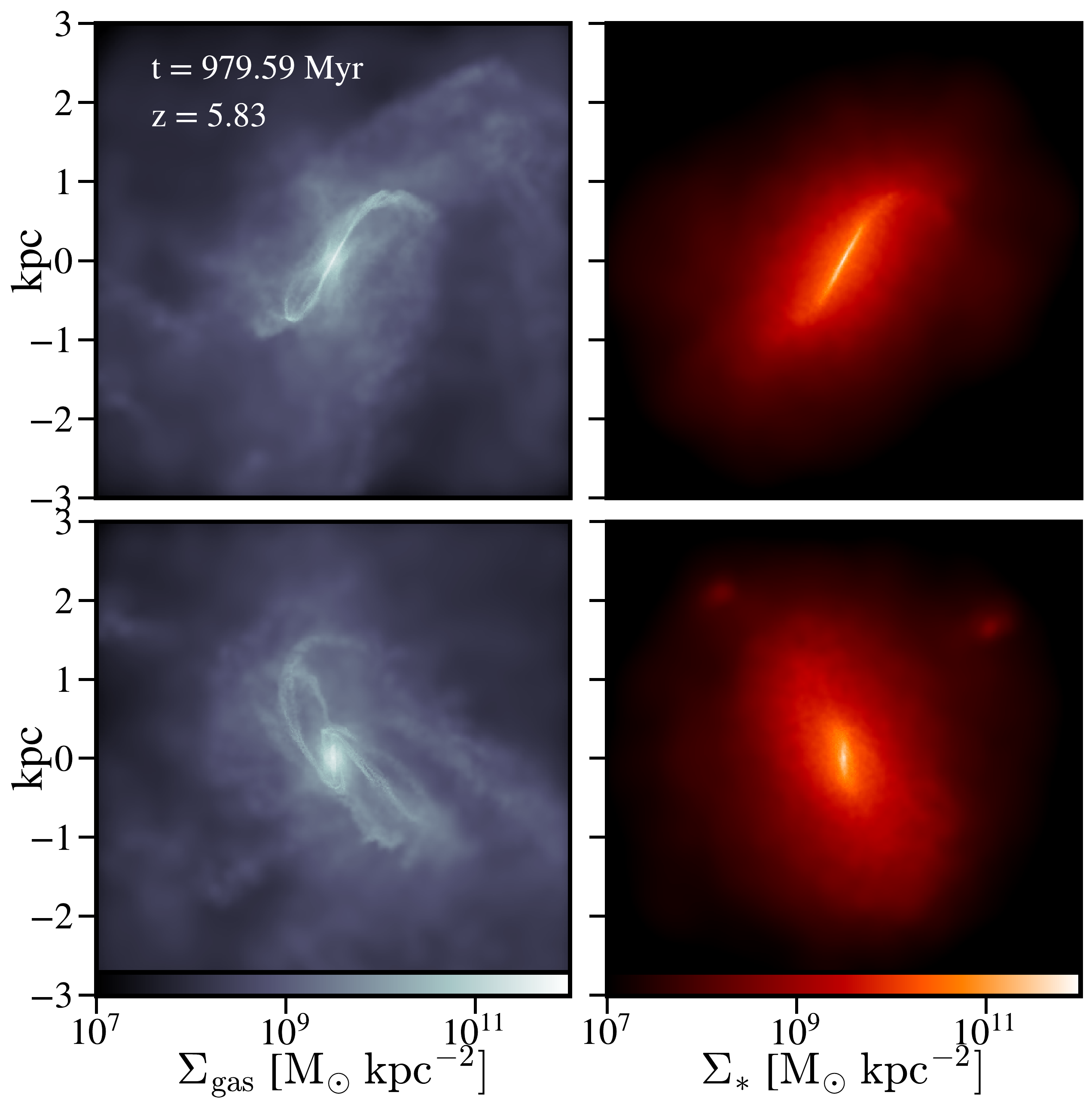}
    \caption{Same as Figure \ref{fig:run0_morphology} but for a different galaxy.}
    \label{fig:run29_morphology}
\end{figure*}

\begin{figure*}
    \centering
    \includegraphics[width=0.48\textwidth]{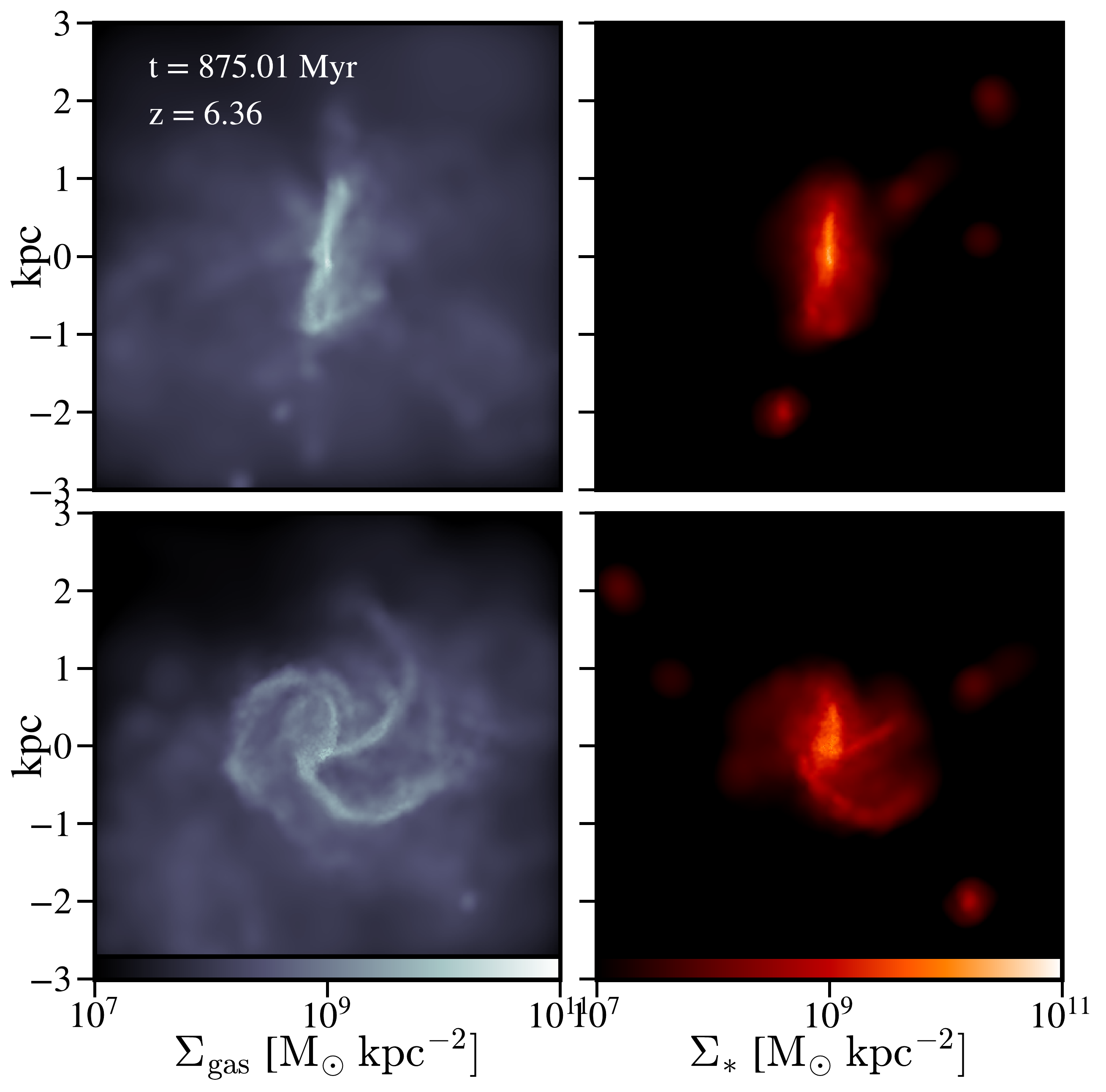}
    \includegraphics[width=0.48\textwidth]{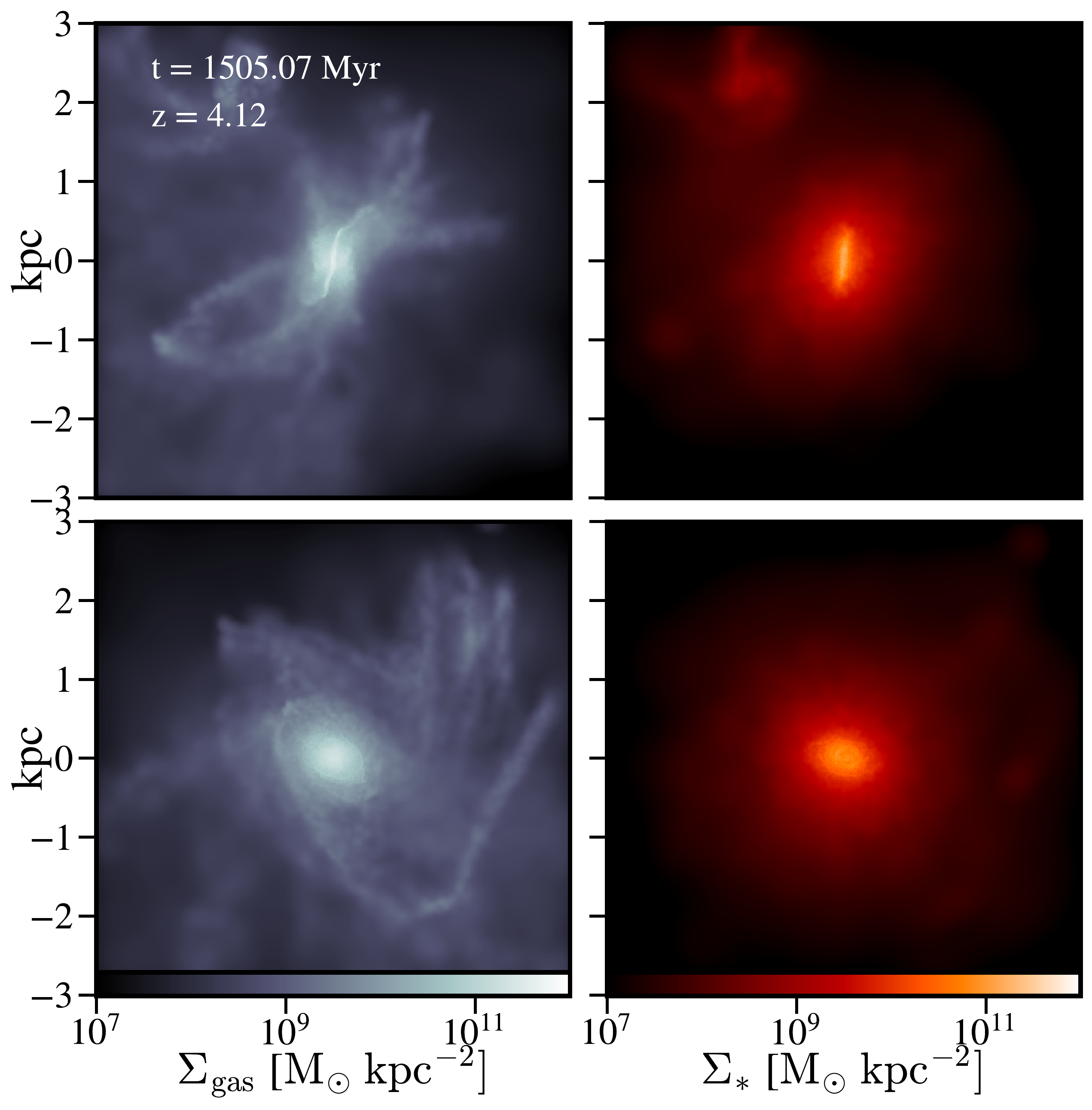}
    \caption{Same as Figure \ref{fig:run0_morphology} but for a different galaxy.}
    \label{fig:run13_morphology}
\end{figure*}

\begin{figure*}
    \centering
    \includegraphics[width=0.48\textwidth]{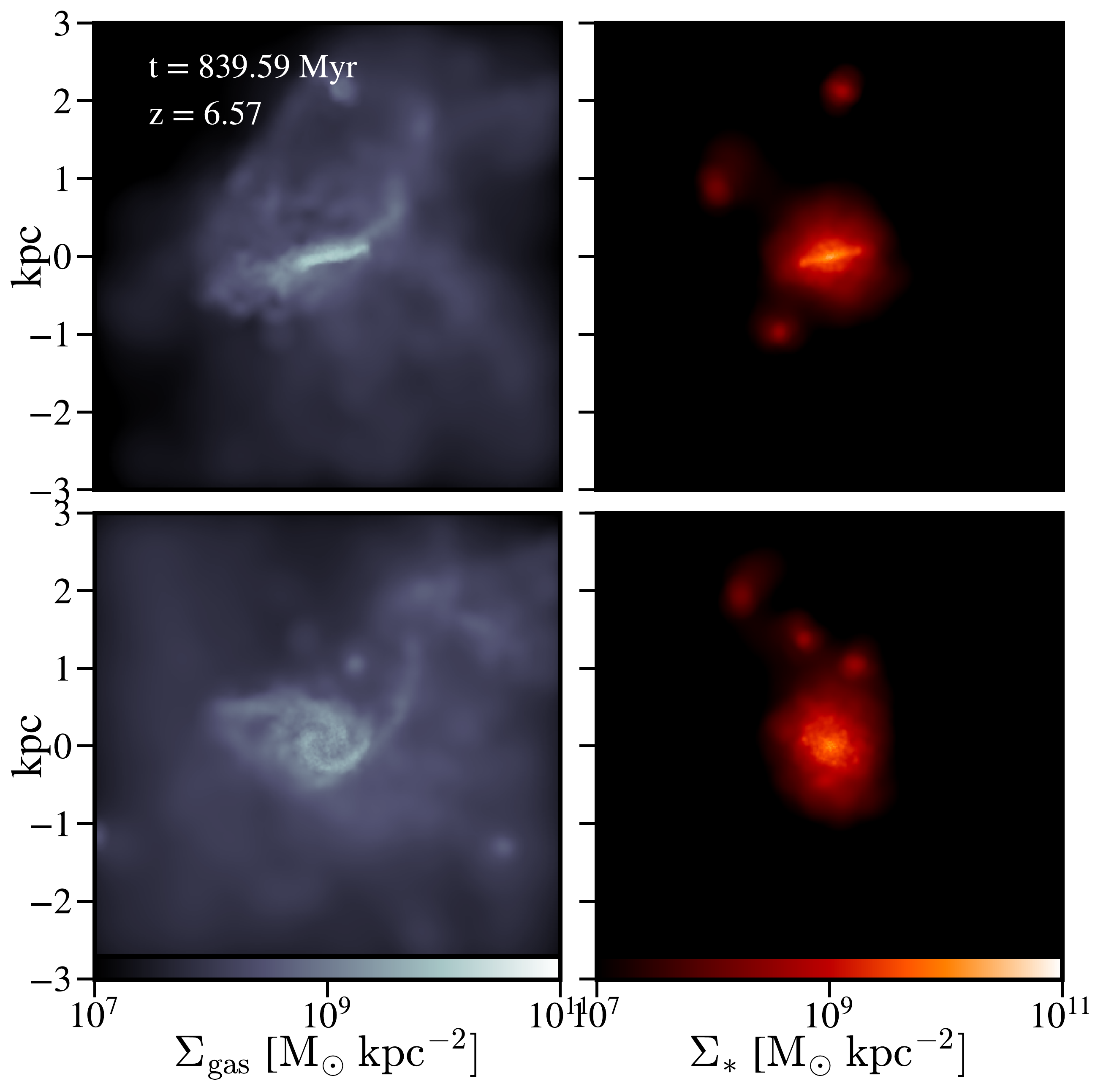}
    \includegraphics[width=0.48\textwidth]{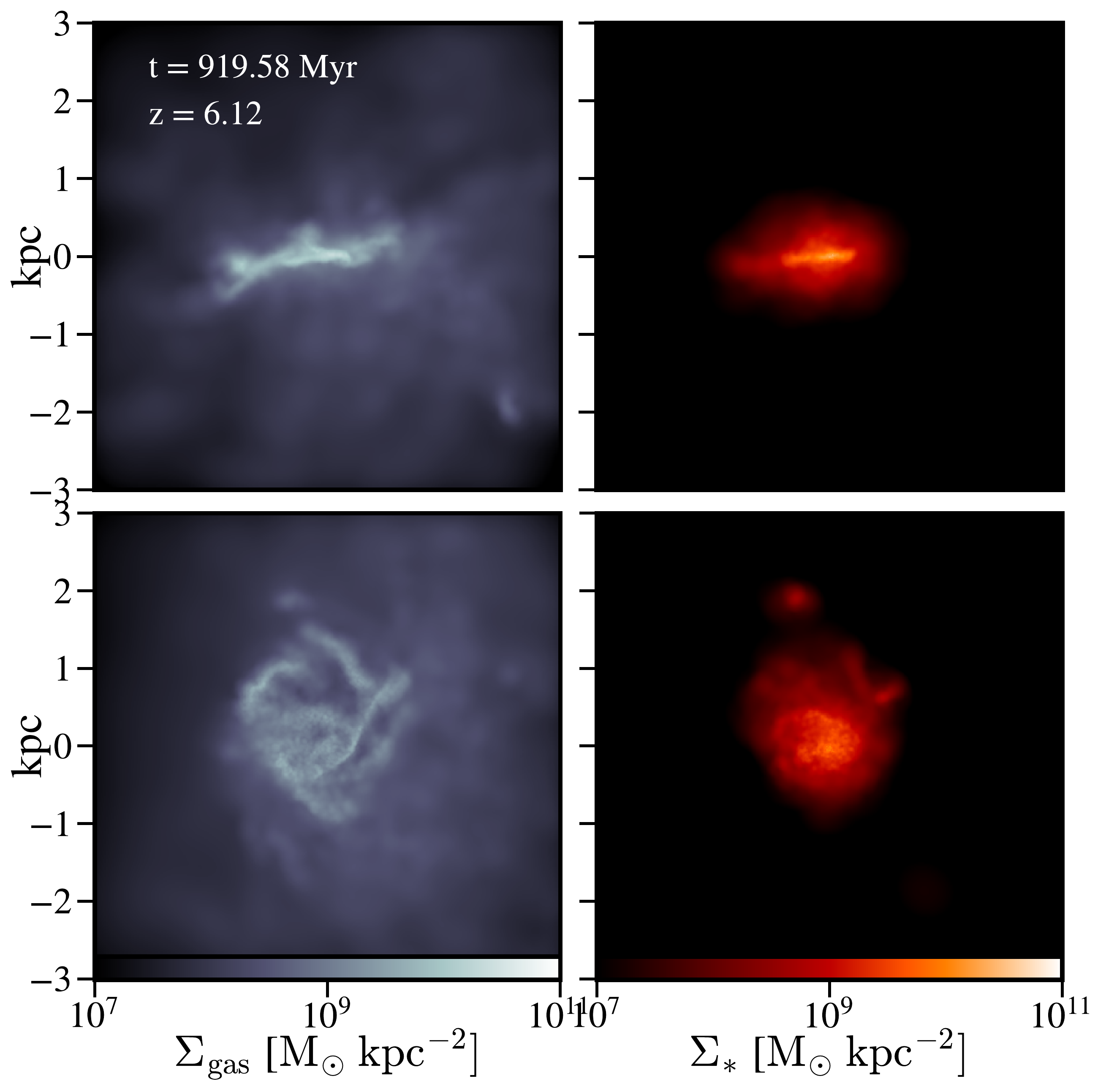}
    \caption{Same as Figure \ref{fig:run0_morphology} but for a different galaxy.}
    \label{fig:run17_morphology}
\end{figure*}

\begin{figure*}
    \centering
    \includegraphics[width=0.32\textwidth]{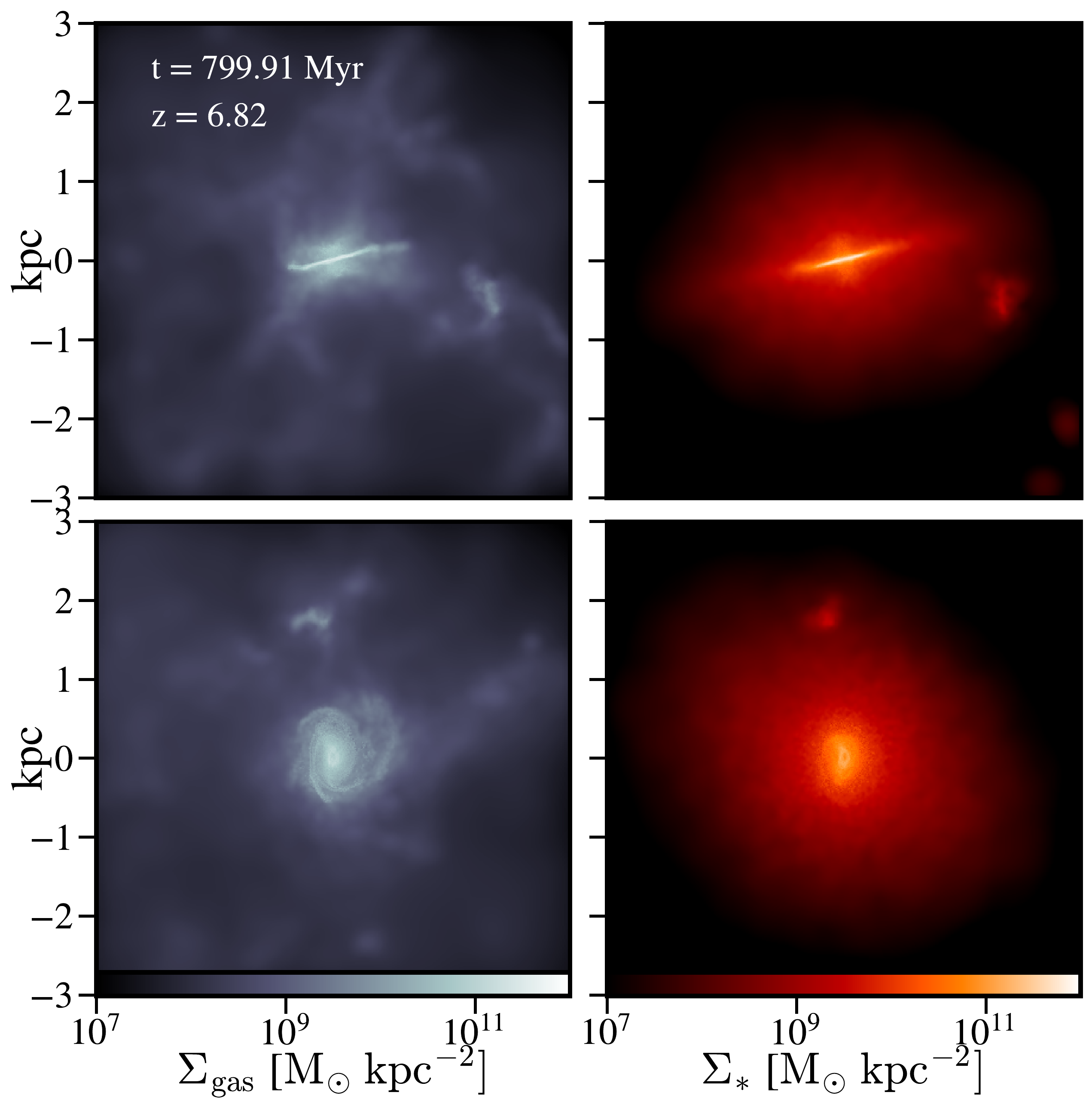}
    \includegraphics[width=0.32\textwidth]{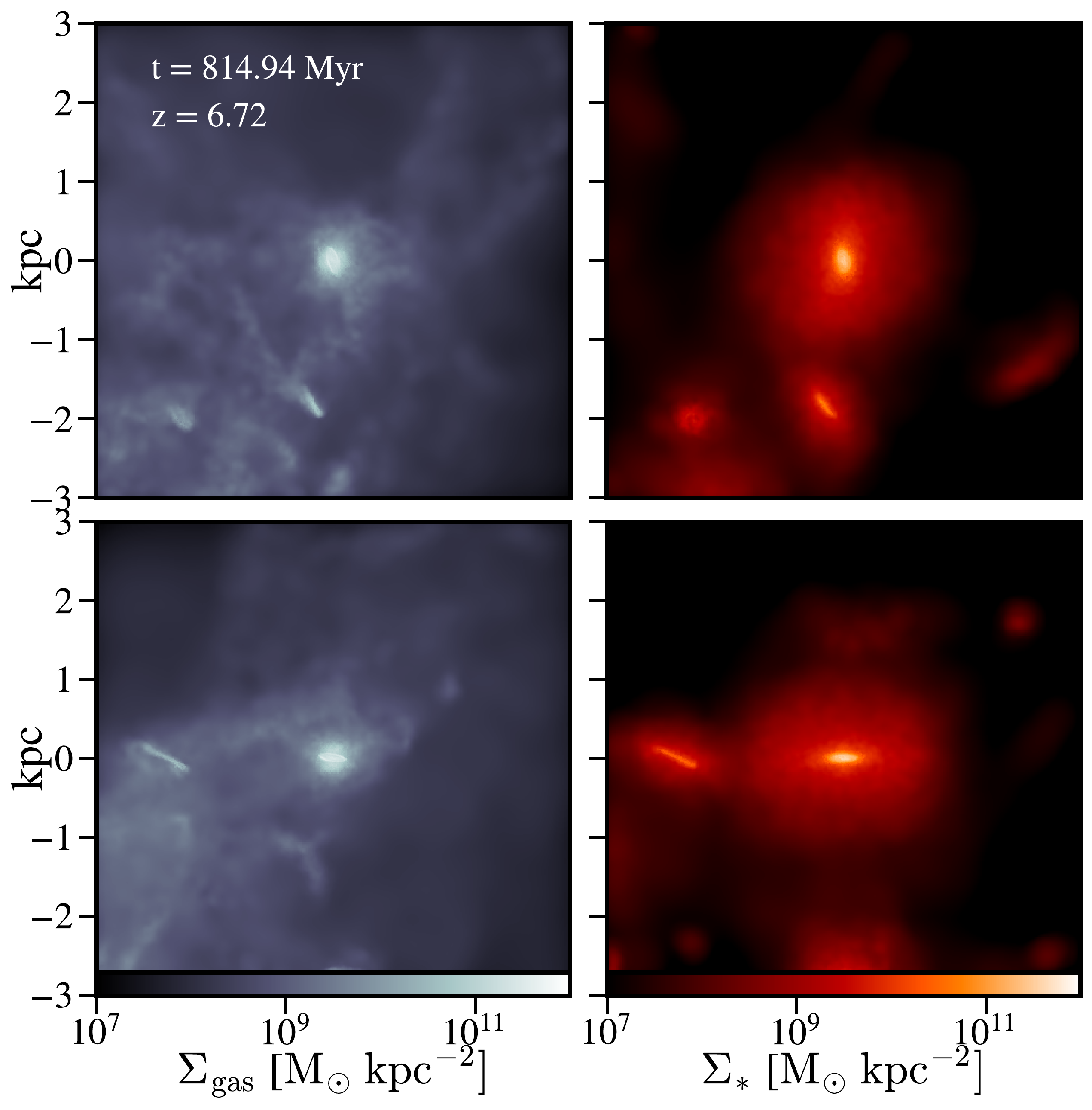}
    \includegraphics[width=0.32\textwidth]{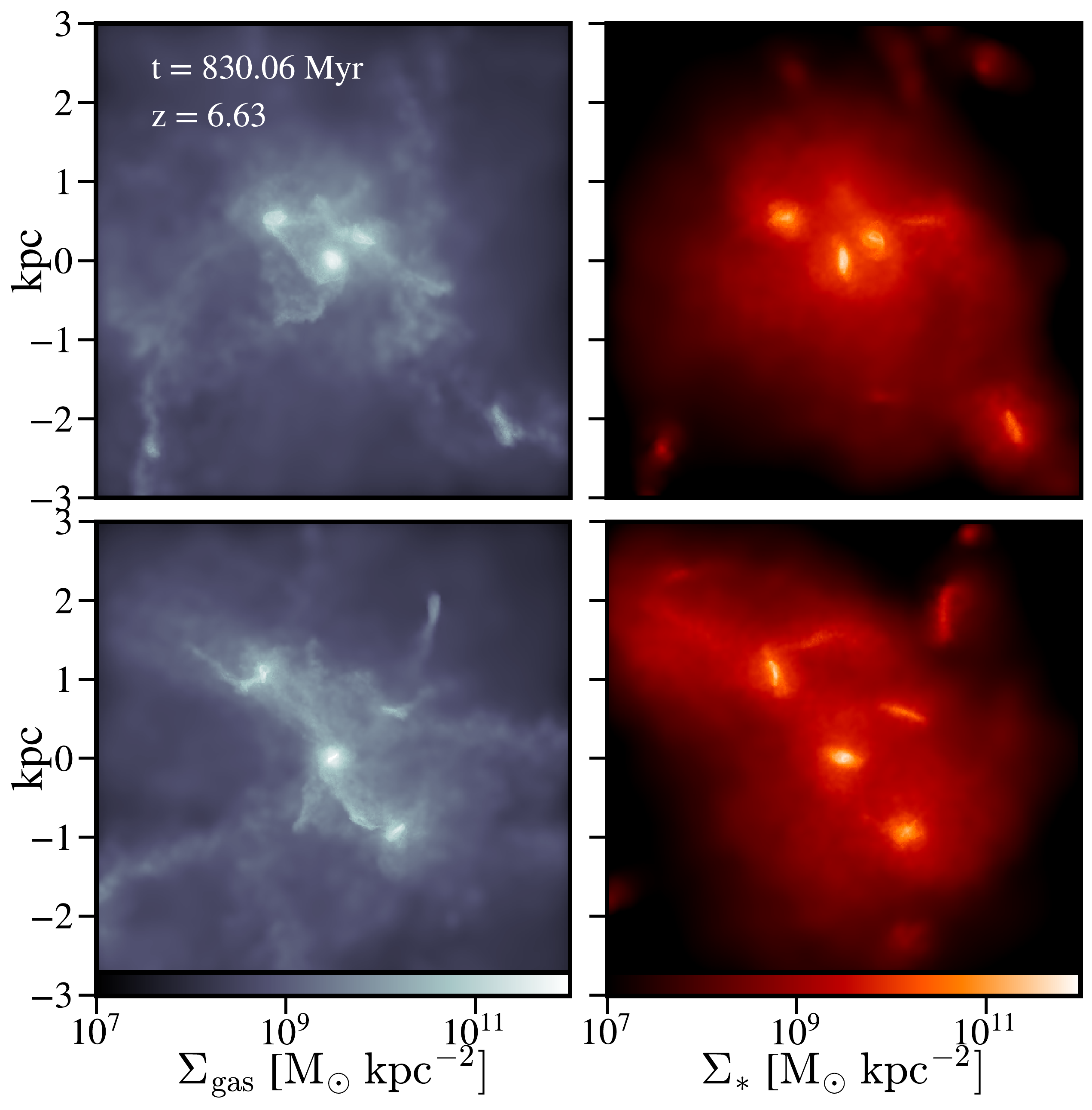}\\
    \includegraphics[width=0.32\textwidth]{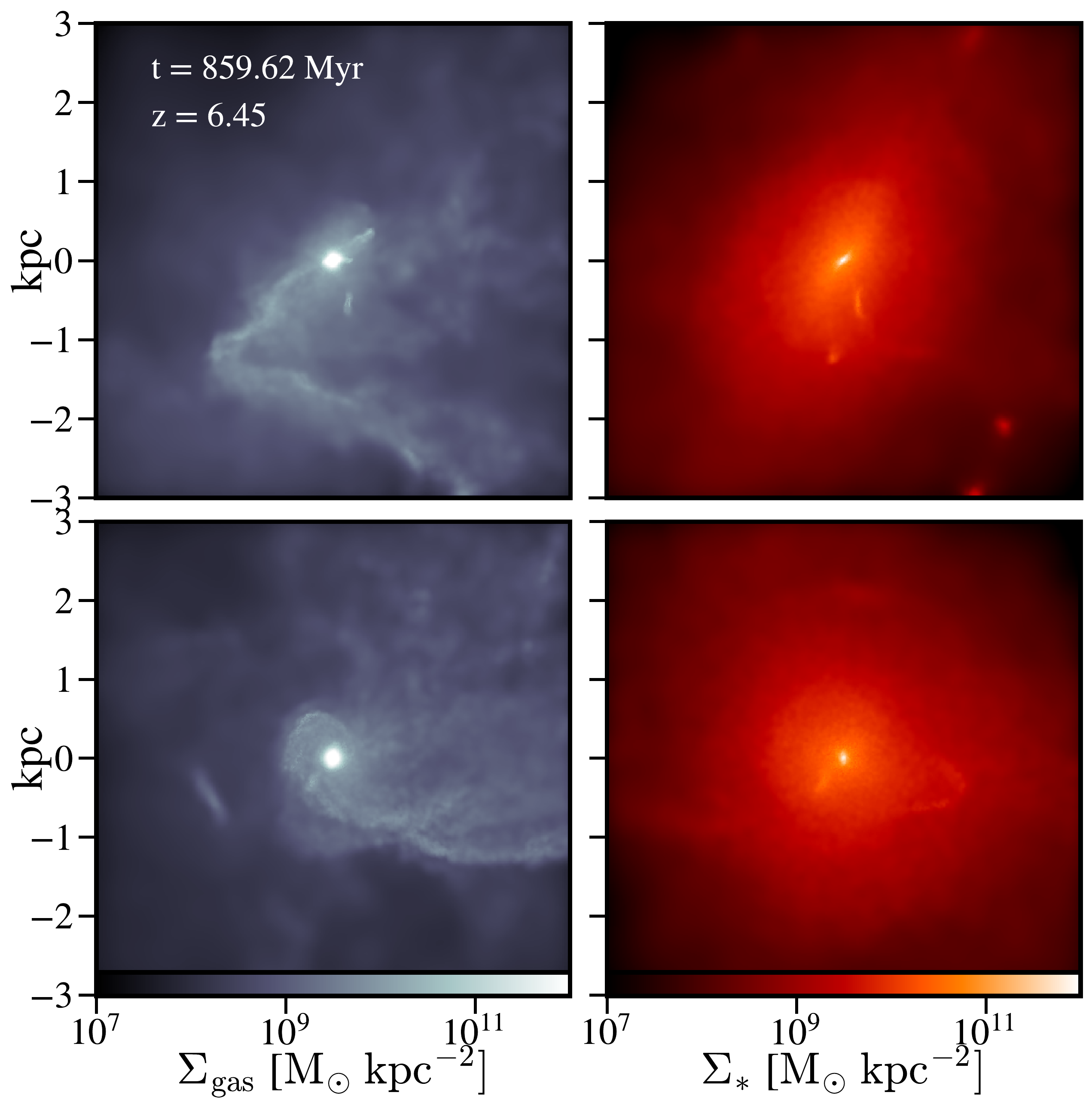}
    \includegraphics[width=0.32\textwidth]{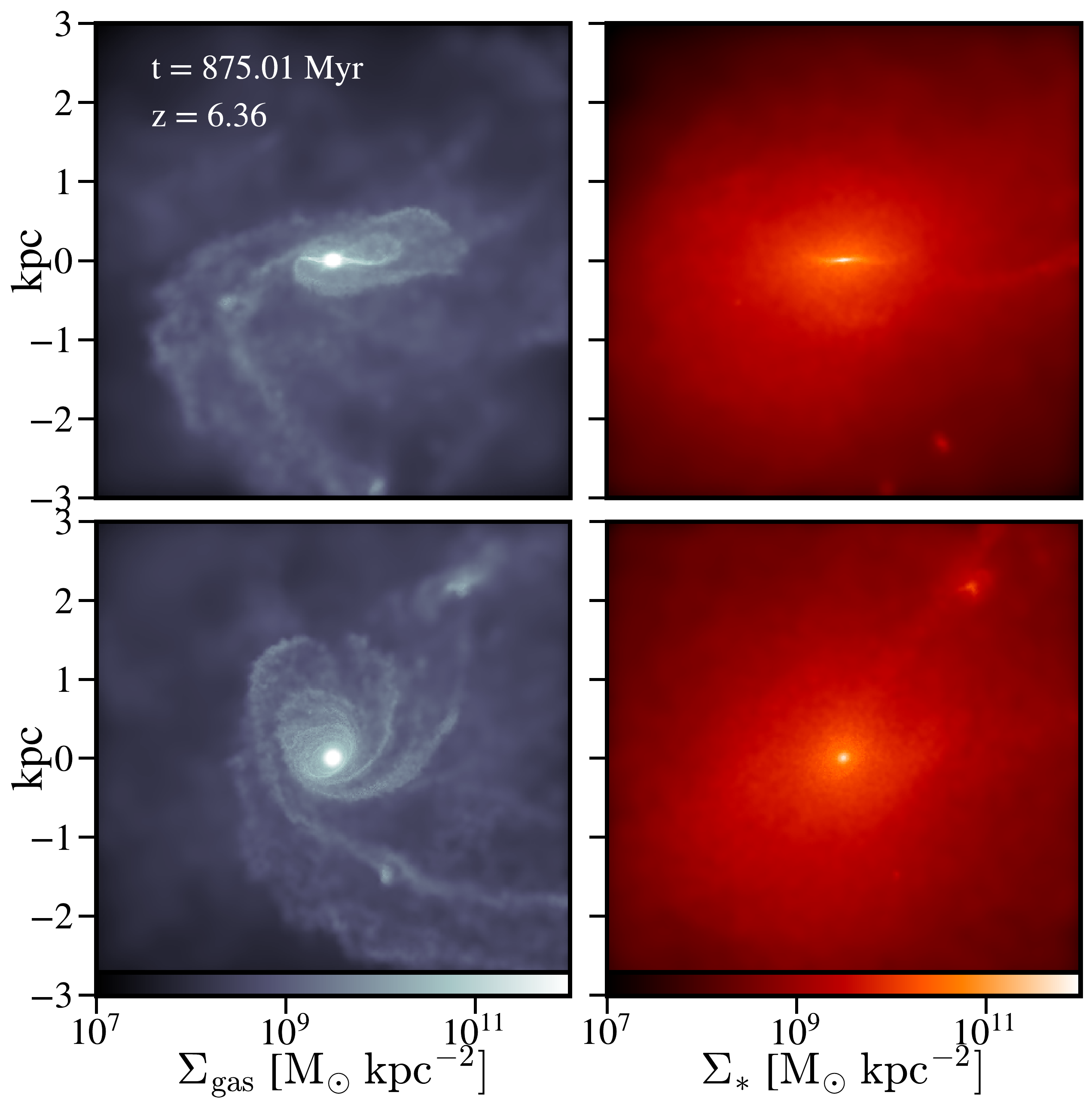}
    \caption{The maximum starburst example undergoing a major merger over 80 Myr.}
    \label{fig:run19_morphology}
\end{figure*}

\bibliography{bib}{}
\bibliographystyle{aasjournal}

\end{document}